# Tidal Coulomb Failure Stresses in the northern Andean intermediate depth seismic clusters: implications for a possible correlation between tides and seismicity

*Gloria A. Moncayo[a], Gaspar Monsalve[b] and Jorge I. Zuluaga[a]*

[a] Solar, Earth and Planetary Physics Group, Computational Physics and Astrophysics Group, Instituto de Física-FCEN, Universidad de Antioquia, Calle 70 No. 52-21, Medellin, Colombia

[b] Departamento de Geociencias y Medioambiente, Facultad de Minas, Universidad Nacional de Colombia, Carrera 80 No. 65-223, Núcleo Robledo, Medellín, Colombia

**Abstract**

A recent statistical analysis of the relationship between tides and seismic activity in Colombia has suggested the existence of correlation anomalies for the case of intermediate depth events in the Bucaramanga nest and the Cauca cluster (Moncayo et al., 2019). In this work, we explore in detail the hypothesis that tides may be triggering seismic activity in these regions and extend the analysis to two other seismic clusters in northern-central South America, specifically in the areas of El Puyo (Ecuador) and Pucallpa (Peru). For this purpose, we use the available focal mechanism information for seismic events at these locations, and calculate for each event the Tidal Coulomb Failure Stress (TCFS) as obtained from estimations of the tidal strain tensor. Tidal strains are computed considering the Earth body tides and the effect of the ocean tidal loading. Since our purpose is to elucidate the role of tides in earthquake nucleation, calculations of the TCFS are conducted not only for the time of earthquake, but also for the time of the closest maximum strain within a window of a few hours before the events. Our results tend to support the hypothesis that tidal stresses are contributing to earthquake generation in all the studied areas; this trend is especially stronger when TCFS are calculated at pre-earthquake times. The physical explanation for the positive contribution of tides to earthquake triggering in intermediate depth clusters may lie in the fact that a wide diversity of fault plane orientations is possible within a relatively small volume of subducted lithosphere, so making the tides more likely to help loosening up the blocks at the fault plane to promote slip.

## 1. Introduction

An earthquake occurs when a sudden release of mechanical energy takes place within the crust, at the hypocenter, where deformations have accumulated over years, decades or centuries. The generation of an earthquake can be described with the so-called seismic cycle (Fedotov, 1965). During one of the phases of this cycle (the interseismic phase), the accumulation of strain brings the system to a critical state that ultimately leads to the occurrence of the earthquake (coseismic phase). Although the origin of the stresses involved in the seismic cycle is mostly endogen (tectonic, volcanic, etc.), we could ask whether, in addition to the endogenous contribution, there could be other phenomena involved in triggering the seismic event. One of these phenomena is the tidal stress, built upon the gravitational interaction of the Earth with the Sun and the Moon. Despite being two or three orders of magnitude smaller than the typical stress drops, tidal strains are associated to stress rates even greater than the tectonic ones (Heaton, 1975; Emter, 1997).

For more than a hundred years, the idea of a relationship between tides and seismicity in the Earth has attracted the attention of scientists around the world. Most recent studies in this topic support a tidal contribution to seismic activity in different areas of the planet (Tanaka et al., 2002, 2004; Tanaka, 2010, 2012; Cochran et al., 2004; Cadicheanu et al., 2007; Metivier et al., 2009; Chen et al., 2012; Tiwari and Chamoli, 2014; Vergos et al., 2015; Xie et al., 2015; Arabelos et al., 2016; Ide et al., 2016); the analysis of moonquakes also reveals a clear tidal contribution (Kolvankar et al., 2010). In some cases, results suggest strong correlations for the case of shallow events of reverse fault type (Tanaka et al., 2002) or shallow thrust earthquakes (Cochran et al., 2004); there is also evidence for tidal triggering of intermediate-depth earthquakes (Cadicheanu et al., 2007). However, there are cases in which results have been weak or contradictory, or do not reveal a link between lunisolar tides and seismicity (Heaton, 1982; Curchin and Pennington, 1987; Hatzell and Heaton, 1989; Vidale et al., 1998; Beeler and Lockner, 2003; Fischer et al., 2006; Ader and Avouac, 2013). Most of the studies up to date have focused on the direct effects of tides on tidal triggering, namely the effect of solid or body tides as measured at or around the hypocenter of the earthquake. Others, however, have also included the so-called indirect effects of the tides (see e.g. Tsuruoka et al., 1995; Vidale et al., 1998; Cochran et al. 2004; Tanaka et al., 2002, 2004, 2006, Tanaka 2010, 2012; Arabelos et al., 2016), namely, the effect that ocean tides could have in tidal triggering.

The indirect effect of the ocean tide is important because near ocean margins, it can be larger than the solid earth tides (e.g. Tanaka et al., 2002). In fact, the load effect in the induced stresses at oceanic basins is an order of magnitude larger than the induced stresses by the solid Earth (Cochran et al., 2004). Therefore, its role as a triggering mechanism should not be overlooked (e.g. Tsuruoka et al., 1995). More interesting is the fact that OTL may disturb the faults by a hold-and-release mechanism, arising from variations in water mass over the ocean basins (Cochran, 2004). This contribution of OTL may result in different effects than those expected by studying only the body tides (BT).

Modeling the OTL is more difficult than modeling solid Earth tides (Agnew, 2007). This fact has delayed the reliable study of these effects on tidal triggering with respect to what have been done with BT. However, recent technological tools and improved geographical models of the ocean basins worldwide, have contributed to improve the estimation of the effect.

In a previous work, we reported the discovery of correlation anomalies (anomalously low p-values of the Schuster Statistical Test) between intermediate depth seismicity in the Bucaramanga nest and Cauca cluster in Colombia (Moncayo et al., 2019) and the diurnal and monthly phase of the body tides. In that study, we neglected the indirect effect of OTL arguing that typical distances of the seismic clusters to the Caribbean and Pacific coasts were relatively large (200-500 km). Since the OTL decreases with distance to the coasts (e.g. Matsumoto et al., 2001; Wendt, 2004), we assumed that considering its effects was not mandatory.

In this work, we revisit the problem and study tidal triggering in the aforementioned seismic clusters, including now the indirect effect of the OTL. Moreover, we take a step further by including in our analysis other two areas of high seismicity rate, close to the West Coast of South America, namely the El Puyo (Ecuador) and Pucallpa (Peru) seismic swarms, where a noticeable concentration of intermediate-depth events has been observed (Zarifi and Havskov, 2003; Soles Valdivia, 2012; Taipe Acosta, 2013). In Moncayo et al. (2019) we focused on devising novel approaches to compute the tidal phases and to perform the statistical analysis of their correlation with seismicity. In that study, we applied our methods to samples of earthquakes with a statistically significant number of events. In this paper, we focus on much smaller samples, namely, those formed by earthquakes whose focal mechanisms are reliably determined. For each earthquake, we compute the local value of the tidal strain, and from them we estimate the so-called Tidal Coulomb Failure Stress (TCFS).

Other authors have attempted to study the TCFS in relation with tidal triggering (e.g. Smith and Sammis, 2002; Cochran et al., 2004; Fischer et al., 2006; Xu et al., 2011; Gallego et al., 2013; Miguelsanz del Alamo, 2016; Bucholc and Steacy, 2016). Thus, for instance, Smith and Sammis (2002) include the tidal Coulomb effective stress response in their analysis of the Northridge and Loma Prieta earthquakes. Independently, Cochran et al. (2004) performed an analysis of worldwide earthquakes considering shear and normal stresses, and found a significant tidal correlation for large tectonic events with increasing tidal stress amplitudes. Fischer et al. (2006) studied the possibility of tidal triggering for an earthquake swarm in the NW-Bohemia/Vogtland region, in Central Europe. In their study they considered only solid Earth tides, due to the large distances between the studied region and the Mediterranean Coast; they reported no significant tidal correlation with the swarms in the region. Xu et al. (2011) conducted a global study of the TCFS with shallow (<60km) and large (M>5) earthquakes for different types of faults. They found a dependence of the TCFS with latitude and fault type. Gallego et al. (2013) analyzed the possible contribution of the solid Earth tides and the OTL to the tremor occurrence around the Chile triple junction region. They found a maximum correlation for Coulomb stresses for faults subparallel to the subducted transform faults. Moreover, they found that the combined effect of both, BT and OTL may facilitate or prevent tremor production in the area. Miguelsanz del Alamo (2016) and Bucholc and Steacy (2016) focused on the very well-known California seismic area. The former proposed new methods to solve the problem of the true nodal plane determination in the focal mechanisms (which is a key problem when analyzing the correlation between the TCFS and tides) and found significant correlations between the maximum TCFS and the origin time of earthquakes; the latter, on the other hand, analyzed both solid Earth tides and OTL and found a correlation between tidal triggering and the magnitude of the tidal stresses.

Earthquake tidal triggering for intermediate-depth earthquakes and/or in seismic nests is not a new research topic (e.g. Curchin and Pennington, 1987; Cadicheanu et al., 2007; 2014). Curchin and Pennington (1987) studied intermediate-depth earthquakes worldwide using only earth tides; they did not find a significant tidal correlation for most intermediate-depth earthquakes. Cadicheanu et al., (2007) considered the fact that the lunisolar attraction can modulate the intermediate depth seismicity and found for the Vrancea zone a significant tidal triggering correlation. In Cadicheanu et al., (2014), the authors searched for some kind of tidal triggering in the three most known seismic nests on Earth: Vrancea (Romania), Bucaramanga (Colombia) and Hindu Kusch (Afghanistan), using a 3D statistical tidal tomography. They found only for Vrancea zone a favorability of tidal triggering, but not for Bucaramanga nor Hindu Kusch. In Bucaramanga they suggested that the time periods between the earthquakes are not associated to tidal triggering (Cadicheanu et al., 2014). They do not use the TCFS criterion in their study. Moreover, and as we argued in our recent work (Moncayo et al., 2019), tidal phases have been improperly computed, especially in the case of the diurnal and monthly phases, in such a way that in most cases, including probably that of Cadicheanu et al. (2014), the strong diurnal and monthly correlations we have observed, have been mostly absent.

This paper is organized as follows: in **Section 2** we start by describing the tectonic setting of the analyzed areas. In **Section 3** we briefly summarize our methods to calculate the direct and indirect effects of tides on Earth for the purpose of our work. **Section 4** describes in detail the calculation of the tidal Coulomb failure stress and the main assumptions behind it. In **Section 5** we present the methods we applied to compute and analyze the TCFS in the studied seismic clusters. **Section 6** presents the results of our computations. **Section 7** is a discussion of the physics involved in the tidal contribution to intermediate depth earthquakes in seismic clusters and the possible tectonic implications. Finally, in **Section 8** we summarize our work and draw the most important conclusions derived from it.

**2. Tectonic setting around the seismic clusters**

The West Coast of South America is characterized by the presence of the Andes mountain range. The Andes Cordillera is the highest non-collisional range worldwide (Ramos, 1999). Several authors distinguish between three different segments in the Andes (see e.g. Gansser, 1973; Ramos, 1999), according to their geological processes and tectonic settings: Northern, Central and Southern Andes. The Northern Block of the Andes (approximately between 12º N and 4ºS) is an interaction zone of three tectonic plates: South America, Nazca and Caribbean, and comprises mostly the territory of Colombia, northern Ecuador, and northwestern Venezuela. The Central Andes (4º- 46º30' S), corresponds to the typical Andean-type orogeny associated to the subduction process (Ramos, 1999), with a flat-slab segment observed in the Peruvian Andes (Ramos, 1999). The Southern Andes (46°30'-52º S) are located south of the triple junction between the Nazca, South America and Antarctic plates (Ramos and Kay, 1992).

In the northern Andean Block and the Central Andes, the region containing the seismic clusters we are interested in (Figure 1), the Nazca plate is subducting under the South American plate, at a relative velocity of about 5.3 cm/yr (Sella et al., 2002, Trenkamp et al., 2002). The subduction of the Nazca Plate in this area seems to be segmented, with a boundary at ~5° N, where flat subduction occurs to the north and a steeper subduction takes place to the south (Vargas and Mann, 2013; Chiarabba et al., 2015; Syracuse et al., 2016). In northern Colombia, the Caribbean plate seems to subduct at a shallow angle at an average rate of 1 - 2 cm/yr (Taboada et al., 1998; Trenkamp et al., 2002). The convergence of these plates, plus the presence of at least two microplates or blocks (Panamá-Chocó and North Andes Blocks), and the existence of several active faults, make the region seismically active and tectonically complex (e.g. Pennington, 1981; Pulido et al., 2003; Cardona et al., 2005).

The flat slab subduction zone in Peru is located between 3ºS and 15ºS and extends over a length of 1500 km, showing the typical absence of volcanism (see e.g. Barazangi and Isaaks, 1976; Gutscher et al., 2000; Tavera and Buforn, 2001; Eakin et al., 2014). Another characteristic of this zone is the presence of the Nazca Ridge between 14ºS and 17ºS, whose subduction can influence the deformation of the mantle in the area (Eakin et al., 2014).

In this tectonic context, we identified 4 regions where an anomalous concentration of intermediate-depth earthquakes takes place (events marked with stars in Figure 1): 1) the Bucaramanga seismic nest (~7ºN, 73ºW), 2) the Cauca seismic cluster (~4.5ºN, 76ºW), 3) the El Puyo seismic cluster (~2ºS, 78W), and 4) the Pucallpa seismic cluster (~8º S, 74º W,).

The Bucaramanga seismic nest corresponds to a concentration of intermediate-depth earthquakes with a mean hypocentral depth of 160 km, which shows the highest concentration of intermediate-depth seismicity worldwide (Prieto et al., 2012). Several authors have tried to explain the origin of this seismic nest (Zarifi and Havskov, 2003; Zarifi et al., 2007; Prieto et al., 2012; Chiarabba et al., 2015); it is still unclear if it is associated with the Nazca Plate (Gutscher et al., 2000; Chiarabba et al., 2015; Syracuse et al., 2016) or the Caribbean Plate (Van der Hilst and Mann, 1994; Sanchez Rojas and Palma, 2014; Yarce et al., 2014; Idarraga-García et al., 2016).

The Cauca seismic cluster is likely related to the subduction of Nazca beneath the North Andean Block (Cortes and Angelier, 2005; Chang et al., 2017). An in-depth analysis of the potential correlation between seismicity and earth tides in those clusters was carried out recently by Moncayo et al. (2019), finding anomalously low values of the Schuster test p-values for the diurnal and monthly tidal components. Those findings motivated this follow-up investigation.

The seismic cluster of El Puyo is located approximately at a distance of 350 km from the trench and is characterized by events with depths between 130 and 220 km and magnitudes that can be greater than Mw=7 (Yepes et al., 2016). Most of the seismic events in the cluster have normal mechanisms (Taipe Acosta, 2013). Zarifi and Havskov (2003) suggested a possible volcanic origin for this cluster, but recent studies show that the El Puyo Cluster could be originated by flexure of the older Farallon slab (Yepes et al., 2016).

The Pucallpa cluster, on the other hand, is described in the literature as a seismic swarm with an extensive regime (Soles Valdivia, 2012). This concentration of events is located at a distance of 600 to 700 km from the trench, and most of the events have depths between 100 and 190 km (Soles Valdivia, 2012). Seismic events in the Pucallpa region show a tendency to be originated in a region parallel to the Cordillera, with a slight deflection to the west (Soles Valdivia, 2012). If this is the case, the Pucallpa seismic cluster could be associated with a possible steepening of the subducting plate (Schneider and Sacks, 1988; Tavera and Buforn, 2001; Soles Valdivia, 2012).

## 3. Calculation of tides and Ocean Tidal Loading (OTL)

Between the two direct effects caused by the gravitational attraction of the Moon and Sun, solid Earth and ocean tides, the former can be described more easily (e.g. Farrell, 1972, 1973; Scherneck, 2001; Zahran et al., 2005; Agnew, 2007). The Ocean Tide Loading (OTL) calculations require the computation of the tidal load numbers (Farrell, 1972). To obtain the OTL, it is necessary to consider a model of the ocean tides and the elastic properties of the Earth (e.g. Zahran et al., 2005; Doan et al., 2006; Agnew, 2007). The effect of ocean loading in a certain location is computed by the convolution of ocean tidal data with Green´s functions (e.g. Farrell, 1972; 1973; Baker, 1984; Jentzsch, 1997; Tanaka et al., 2002; Tanaka, 2010, 2012; Doan et al., 2006; Penna et al., 2008). The OTL displacement $u(\vec{r})$ at a given point can be written in the following form (Farrell, 1973):

$$u(\vec{r}) = \int_\Omega \rho G(|\vec{r} - \vec{r}'|) H(\vec{r}') d\Omega \qquad [1]$$

where $\rho$ is the density of sea water, $G$ represents the Green´s function, which depends on the distance between the position vectors on the Earth's surface $\vec{r}$ and $\vec{r}'$ (Farrell, 1973), and $H$ corresponds to the tidal amplitude at $\vec{r}'$. The integral is calculated over the total area of water described by $\Omega$.

Considering that the response of the ocean to the tidal forces is not uniform, both phenomena, solid Earth tides and OTL, are generally not in phase (Jentzsch, 1997; Wilcock, 2009). Another aspect to consider is that the local properties of the crust and mantle affect the OTL, whereas the solid or body tides depend on the global properties of the Earth (Farrell, 1972).

The OTL effect decreases with distance to the seashore (e.g. Doan et al., 2006). Several studies highlight the role that the OTL plays in the search for a correlation between tides and seismicity, which is a factor that cannot be neglected, especially if the seismic events are in areas near the coasts (e.g. Tsuruoka et al., 1995; Jentzsch et al., 2001; Wilcock, 2001, 2009; Cochran et al., 2004; Tanaka et al., 2002, 2006; Tanaka, 2010, 2012; Thomas et al., 2012). As mentioned by Farrell (1972), an important characteristic of the OTL is its much more irregular distribution, in comparison with the terrestrial tide, which shows smooth variations at the Earth´s surface. A detailed theoretical description of ocean tides can be found in e.g. Melchior (1966), Jentzsch (1986, 1997), Zahel (1997), and Agnew (2007).

In Figure 2 we show the components of the strain and radial displacement produced by the OTL effect in the northwestern regions of South America around Colombia and for different major tidal components. As expected, the largest effects are close to the west coast, in the pacific basin for the M2 and S2 components, and around the Caribbean basin for the O1 constituent. Although the radial displacement (~10 mm) is small as compared to the typical total tidal displacement (~200 mm, see Figure 6 in Moncayo et al. 2019), for events at or close to the ocean basins, the OTL strains are comparable to those produced by the solid tides.

## 4. Tidal triggering and the tidal Coulomb failure stress (TCFS)

The feasibility that tides trigger seismic activity in a given region of the Earth strongly depends on the comparison of the magnitude of the local endogenic (e.g. tectonic) and the external tidal effects. The stresses induced by Earth tides are of the order of $10^3$ Pa while those of the OTL can reach values as large as $10^4$ Pa (e.g. Cochran et al., 2004). Therefore, tidal stresses are 3-4 order of magnitudes smaller than those of tectonic origin, which are of the order of $10^5$-$10^7$ Pa (Vidale et al., 1998; Metivier et al., 2009). This comparison is one of the main drawbacks of the hypothesis of tidal triggering of seismicity.

However, tidal stress rates can be comparable and even much larger than tectonic stress rates (e.g. Emter, 1997). Assuming, for instance, that OTL stresses build up during half of the semidiurnal cycle, tidal stress rates can be as large as ~$1.7 \times 10^3$ Pa/h (near the coast, the tidal loading signal can overlap the solid tide signal, e.g. Jentzsch, 1997). This value is two orders of magnitude larger than tectonic stress rates ~17 Pa/h (Heaton, 1975). Therefore, and in contrast to tectonic stress, tidal stress may contribute to activate faults already in a critical state (e.g. Smith and Sammis, 2002; Bucholc and Steacy, 2016). A similar argument has been raised for the case of volcanic regions (Jentzsch et al., 2001).

Estimating the value of the tidal stresses that act in a fault during or before an earthquake is not a trivial matter (Stein, 1999; Vidale et al., 1998; Cochran et al., 2004; Fischer et al., 2006; Wilcock, 2009; Xu et al., 2011; Miguelsanz del Álamo, 2016; Bucholc and Steacy, 2016). In this work we will use the typical approach used by many authors, namely estimating the tidal stress as a linear combination of the normal and shear stress at the fault plane.

This "combined stress" is called the Tidal Coulomb Failure Stress (TCFS):

$$\sigma_c = \sigma_s + \mu_f \sigma_n \qquad [2]$$

Here $\mu_f$ is the apparent friction coefficient (King et al., 1994), $\sigma_n$ is the normal tidal stress, which is considered positive for extension, and $\sigma_s$ is the tidal shear stress, positive in the slip direction of the fault. Positive values of the TCFS components would be normally associated to fault slip (e.g. Stein, 1999; Gallovic et al., 2008; Miguelsanz del Alamo, 2016).

There is no consensus on the best value for $\mu_f$. It can be as low as 0.2 and as large 0.8. Several authors, however, have found that a value $\mu_f = 0.4$ leads to the largest values of the TCFS when a correlation between tides and seismicity seems to exist (e.g. Cochran et al., 2004; Fischer et al., 2006; Bucholc and Steacy, 2016; Miguelsanz del Alamo, 2016).

The value of normal and shear stresses can be computed from the tidal traction vector, $\boldsymbol{T}$, whose components are defined in terms of the stress tensor $\sigma_{ij}$ as $\boldsymbol{T_i} = \sigma_{ij} n_j$, where $\boldsymbol{n}$ is a vector normal to the fault plane. Using the traction vector, the normal ($\sigma_n$) and shear ($\sigma_s$) components of the stress are (Xu et al., 2011; Miguelsanz del Alamo, 2016):

$$\sigma_s = \boldsymbol{T} \cdot \boldsymbol{s} \qquad [3]$$
$$\sigma_n = \boldsymbol{T} \cdot \boldsymbol{n} \qquad [4]$$

where we have additionally introduced the slip vector $\boldsymbol{s}$.

The normal and slip components are given in terms of the parameters of the fault mechanisms, namely the strike ($\varphi$), dip ($\delta$) and rake ($\lambda$), by (Miguelsanz del Alamo, 2016):

$$s = \begin{pmatrix} \sin\varphi \cos\lambda - \sin\lambda \cos\delta \cos\varphi \\ \cos\lambda \cos\varphi + \sin\lambda \cos\delta \sin\varphi \\ \sin\lambda \sin\delta \end{pmatrix} \quad [5]$$

$$n = \begin{pmatrix} \cos\varphi \sin\delta \\ -\sin\varphi \sin\delta \\ \cos\delta \end{pmatrix} \quad [6]$$

The constitutive equation for an isotropic material, provides us the relationship between the stress and the components of the strain $e_{ij}$:

$$\sigma_{ii} = \lambda e_{kk} + 2\mathcal{G}e_{ii} \quad [7]$$

$$\sigma_{ij} = 2\mathcal{G}e_{ij} \quad [8]$$

Here, $\lambda$ and $\mathcal{G}$ are the Lamé coefficients, with $\mathcal{G}$ the *rigidity* of the material, which for the Earth's crust has values between 30 GPa and 75 GPa (Turcotte and Schubert, 2002). On the other hand, the Lamé coefficient $\lambda$ is obtained from rigidity $\mathcal{G}$ and Poisson's ratio $v$ (Stein and Wysession, 2003; Miguelsanz del Alamo, 2016)

$$\lambda = \frac{2\mathcal{G}v}{1-2v} \quad [9]$$

Combining Eqs. (5-9), the traction vector can be written explicitly as:

$$\boldsymbol{T} = \boldsymbol{\sigma} \cdot \boldsymbol{n} =$$
$$\frac{2\mathcal{G}v}{1-2v} \begin{bmatrix} ((1-v)e_{xx} + v(e_{yy} + e_{zz}))(\cos\varphi \sin\delta) - (1-2v)e_{xy} \sin\varphi \sin\delta (1-2v)e_{xz} \cos\delta \\ (1-2v)e_{xy}(\cos\varphi \sin\delta) - (v(e_{xx} + e_{zz}) + e_{yy}(1-v)) \sin\varphi \sin\delta + (1-2v)e_{yz} \cos\delta \\ (1-2v)e_{zx}(\cos\varphi \sin\delta) - (1-v)e_{yz} \sin\varphi \sin\delta + ((1-v)e_{zz} + v(e_{xx} + e_{yy})) \cos\delta \end{bmatrix} \quad [11]$$

here, the strain components $e_{xx}, e_{yy}, e_{zz}, e_{xy}, e_{yz}, e_{xz}$ are expressed using a local Cartesian coordinate system ($x$ corresponds to east E, $y$ to north N and $z$ to the up-direction U) (e.g. Smith and Sammis, 2002; Fischer et al., 2006; Jaeger et al., 2007; Miguelsanz del Alamo, 2016). Explicitly these components are:

$$e_{xx} = \frac{\partial u_x}{\partial x} = e_{EW} \qquad \text{Horizontal strain 90°} \qquad [12]$$

$$e_{yy} = \frac{\partial u_y}{\partial y} = e_{NS} \qquad \text{Horizontal strain 0°} \qquad [13]$$

$$e_{zz} = \frac{\partial u_z}{\partial z} = e_{UU} \qquad \text{Vertical strain} \qquad [14]$$

$$e_{xy} = e_{yx} = e_{NE} - \frac{1}{2}(e_{NS} + e_{EW}) \text{ (horizontal tidal shear strain)} \quad [15]$$

$$e_{xz} = e_{zx} = e_{EZ} - \frac{1}{2}(e_{EW} + e_{UU}) \qquad [16]$$

$$e_{yz} = e_{zy} = e_{NZ} - \frac{1}{2}(e_{NS} + e_{UU}) \qquad [17]$$

In the equations above, $e_{NE}$, $e_{EZ}$ and $e_{NZ}$ are the diagonal strains in the N45E, E45Z and N45Z directions, respectively.

For earthquakes close to Earth's surface we can assume a free surface boundary condition, i.e. the shear strains $e_{xz}$ and $e_{yz}$ almost vanish (Harrison, 1976). This implies $\sigma_{xz} = \sigma_{yz} = 0$ (Zürn and Wilhelm, 1984). Additionally, we can also assume a state of plane strain, namely $e_{zz} = 0$. Both of these assumptions are common when studying shallow earthquakes (e.g. Wilcock, 2009; Smith and Sammis, 2002; Fischer et al., 2006). In this case, the traction vector is finally obtained as:

$$T = \frac{2Gv}{1-2v}\begin{bmatrix} (1-v)e_{xx} + v(e_{yy})(\cos\varphi\sin\delta) - e_{xy}\sin\varphi\sin\delta \\ (1-2v)e_{xy}(\cos\varphi\sin\delta) - (ve_{xx} + e_{yy}(1-v))\sin\varphi\sin\delta \\ v(e_{xx} + e_{yy})\cos\delta \end{bmatrix} \quad [18]$$

Although for the calculation of the tidal stress we need to consider the earthquake depth (Tsuruoka et al., 1995), for the case of intermediate-depth earthquakes the previous simplifications are still valid. Shear stresses are near zero from Earth´s surface to approximately 200 - 300 km depth (Varga and Grafarend, 1996, 2017).

Depending on the sign of the involved stress, the slip of the fault may be favored or not (e.g. Tusuroka et al., 1995). A positive normal stress means the total normal pressure on the fault decreases and so the slip will be favored (e.g. Miguelsanz del Alamo, 2016; Bucholc and Steacy, 2016). The same happens for the shear stress; if it is positive, it implies that the movement of the fault is favored (Xu et al., 2011).

## 5. TCFS for intermediate depth events in Northwestern South America

In Moncayo et al. (2019), we presented statistical evidence of a positive correlation between tides and seismicity for the Bucaramanga seismic nest and the Cauca cluster. To actually demonstrate that these correlation anomalies correspond to a causal effect (tidal triggering), we need to go further (see e.g. Ader and Avouac, 2013). Here, we will study the problem from a geophysical point of view, computing the tidal stresses acting on several of the earthquake nucleation areas identified in Northwestern South America.

As explained in previous sections, this goal requires the estimation of the components of the strain tensor at the place and time of seismic events, besides knowing the geometrical characteristics of the faults where those events arise. The main inputs for the present study consist of information about seismic events occurred in northwestern South America and time series of BT and OTL at the place and time of the selected events.

### 5.1. The selected dataset

For the seismic events within the Bucaramanga nest and the Cauca cluster, we used the datasets from the Colombian National Seismological Network (RSNC), managed by the Colombian Geological Survey and freely available online at http://seisan.sgc.gov.co/RSNC. The dataset contains 167162 earthquakes recorded from 1993 to 2017. Approximately 70% of the events in the database correspond to intermediate-depth seismic events occurred in the Bucaramanga nest and Cauca cluster (Moncayo et al., 2019). The number of events with information about moment tensors and focal mechanisms is however restricted.

We extracted source mechanism information mainly from Cortes and Angelier (2005), which contains events between 1964 and 2002, and from the dataset of the Global Centroid-Moment-Tensor (CMT) catalog from 1997 to 2017 (Dziewonski et al., 1981; Ekström et al., 2012). This information freely available online at www.globalcmt.org. Additional information was obtained from Salcedo et al. (2001) and Tabares et al. (1999), both of which contain earthquakes from 1966 to 1992. Chang et al., (2019) provided focal mechanism information for events from 2010 through 2014 in the Cauca cluster. The focal mechanism information of the seismicity in the El Puyo and Pucallpa regions was all obtained from the CMT catalog. Considering this combined data set, we must assume a wide range of uncertainties in the nodal planes attitude in the focal mechanisms, which we take to be between 10 and 45 degrees.

Using the preceding criteria, we selected 56 earthquakes from the Bucaramanga nest and 89 from the Cauca cluster. For the Bucaramanga nest, the selected earthquakes range in magnitude between 4.2 and 6.2 and in depth between 120 and 172 km; in the Cauca cluster, the magnitudes of the selected events go from 2.5 to 6.3 and hypocentral depths are between 60 and 189 km (Supplementary Tables S1 to S4). The focal mechanisms of the Bucaramanga nest have a wide variability (Zarifi and Havskov, 2003; Frohlich and Nakamura, 2009), but most of them correspond to reverse mechanisms (Frohlich and Nakamura, 2009). The Cauca cluster also shows some variability in the focal mechanism information, but the strike slip fault type predominates (Chang et al., 2019).

Additionally, from the CMT catalog we extracted focal mechanisms for 23 earthquakes in the El Puyo seismic area, and 28 for the Pucallpa seismic cluster. These earthquakes have magnitudes between 4.8 to 7.1 and depths between 149 and 198 km for El Puyo, and for Pucallpa magnitudes range between 4.9 and 6.6, and depths between 144 and 197 km (Supplementary Tables S5 to S8).

Following Cadicheanu et al. (2014), we considered all main shock and aftershocks of the intermediate-depth earthquakes in our selected sample. In our previous study (Moncayo et al., 2019), the whole earthquake database of Colombia was declusterized. However, for the present analysis, the aftershocks are not filtered out. The possibility of cascading triggering phenomena in case of a not declusterized database should then be considered. In our study, the database of intermediate-depth earthquakes with information of focal mechanism was not declusterized. For seismic events in the Bucaramanga nest (56 events from 1977 to 2016, median local magnitude ML= 4.9), we can observe that the events are distributed in a broad range of time; 84% of the events are separated by more than 10 days, and even most of them occurred with a separation of more than 100 days; only 7% of the events where separated by a time interval less than 10 days. In the Cauca cluster (89 events from 1966 to 2015, median local magnitude ML= 5.5), 78% of the events are separated by times larger than 10 days. In El Puyo cluster (23 events, median local magnitude ML=5.4), 95% of the events are separated by a time interval larger than 10 days. All the studied events from the Pucallpa cluster (28 events, local median magnitude ML= 5.6) are separated by time intervals larger than 10 days. This gives us confidence that our data are mostly free of aftershocks. In addition, in the case of the Bucaramanga nest, the productivity of the aftershocks has not been reliably detected. Prieto et al. (2012) suggest that the possible reason for the limited aftershock detection is the background seismicity, because it acts as a masking effect, hiding the aftershocks.

**5.2. OTL and BT time series computation**

Considering the importance of including the OTL effect in the exploration of a relationship between tides and seismicity, we chose the program GOTIC2 (Matsumoto et al., 2001) to obtain the time series of the indirect effect (OTL) and the joint effect of the solid Earth tides (body tides or BT) and OTL. GOTIC2 (Matsumoto et al., 2001) is a FORTRAN program, which computes and predicts theoretical BT and OTL in the time domain.

Using GOTIC2 we obtained tidal time series for BT, OTL and the combined effect in an interval of 30 days before and 30 days after the earthquake. The time series were computed considering the models NAO.99b (global ocean tide model with $0.5º$ resolution, Matsumoto et al., 2001) and the long period model NAO.99L (Takanezawa et al., 2001; Matsumoto et al., 2001). Both models include 21 tidal constituents: 16 constituents with short period in the diurnal and semidiurnal band (M2, S2, K1, O1, N2, P1, K2, Q1, M1, J1, OO1, 2N2, Mu2, Nu2, L2, T2) and 5 long period constituents (Mtm, Mf, Mm, Ssa, Sa).

Although near the coasts it is advisable the usage of combined ocean models of global and regional character (Matsumoto et al., 2001), in our study we use only the global model because most of the analyzed seismic events occurred at a distance larger than 200 km from the coast, where the resolution of the model is enough. A possible improvement of our model will otherwise consider coastal geometry (Jentzsch et al., 2000).

The theoretical OTL effect for the major tidal constituents M2 (principal lunar semidiurnal), S2 (principal solar semidiurnal) and O1 (principal lunar diurnal) is shown for Colombian territory and neighboring regions in Figure 2. The M2 constituent shows the largest amplitude variations, both in the OTL and BT (e.g. Penna et al., 2008). The radial displacement shows its highest values in the Pacific Coast of Colombia for both the semidiurnal principal short period components M2 and S2. The Moon diurnal principal component O1 shows the highest values at the Caribbean Coast. Between M2 and S2 there are differences of one order of magnitude, and between M2 and O1, in some cases the difference reaches two orders of magnitude. If we consider the effect of the OTL on strain, the Moon principal semidiurnal component shows that the tidal effect concentrates in the area of the Pacific Coast for the N-S and E-W components. In all cases the strain decreases towards the continent, becoming practically negligible at distances larger than 300 km.

**5.3. TCFS computation**

To compute the TCFS, we calculated for all the selected events, time series of the longitudinal tidal strains for both BT and OTL, as well as the combined effect. Our time series include information about the strain in the north-south, east-west and vertical directions, and in the diagonal direction N45°E. The normal strain components $e_{yy}$ in north-south direction (also called $e_{NS}$), $e_{xx}$ in east-west direction (known also as $e_{EW}$), the shear component $e_{xy}$ and the $e_{zz}$ component were also calculated at the reported hypocenter of the event and around the origin time.

In all cases, independently of earthquake depth, for computing the tidal stresses we use the criteria of Varga and Grafarend (2017), according to which for depths less than 200-300 km the stress tensor components $\sigma_{zz}$, $\sigma_{xz}$, $\sigma_{yz}$, $\sigma_{xy}$ are negligible. Additionally we use Green's functions from GOTIC2 that were calculated for the surface of the Earth and verified that strains do not differ significantly from those computed with ETERNA 3.40 (Wenzel, 1996, 1997), which is better suited to compute tidal strains at any depth.

The traction vector, the normal and tangential stresses on the fault plane, were computed using the Cauchy's formula. The latter two would significantly affect fault rupture (e.g. Tanaka et al., 2002; Xie et al., 2015). In all cases, tidal stresses were computed at hypocenter depth, as suggested by Tsuruoka et al. (1995). The normal and tangential stresses were used to obtain the TCFS of the earthquakes according to Eq. (2).

6. **Results**

The normal and shear stresses computed for all the events in our selected set of earthquakes are presented in Tables S1 and S2 of the supplementary material. From these values, we compute the TCFS using Eq. (2) for at least two different values of the friction coefficient, namely $\mu_f = 0.2$ and $\mu_f = 0.4$, and a rigidity of 75 GPa, which can be representative of the upper mantle (Turcotte and Schubert, 2002) and appropriate for intermediate depth earthquakes. Since from the information provided by the focal mechanisms it is difficult to establish the true nodal plane, we have computed the TCFS for both nodal planes. For each event, these two values of TCFS are considered as independent events when computing all the statistics presented below.

In Figures 3-6 we show histograms of the resulting values of the TCFS (considering both BT and OTL contributions), as calculated for the events in our dataset. In order to statistically characterize the estimated TCFS distributions, we compute several central and non-central momenta of the sample of TCFS: the arithmetic average (estimated mean), the square-root of the sample variance (estimated standard deviation), estimated skewness and estimated kurtosis of the distribution.

A positive causal relationship can be suggested if the estimated TCFS mean is positive, with an estimated standard deviation smaller than or at least of the same order of magnitude of the mean. A large positive skewness is also suggestive of a distribution biased towards positive TCFS values. Moreover, a large positive kurtosis may also suggest an underlying distribution which is heavy tailed. In summary, a combination of a very positive mean, a relatively small standard deviation, and positive and large values of the skewness and kurtosis, supports the hypothesis of tidal triggering.

Computations of the TCFS were performed at two different times: 1) at the precise time of the earthquake occurrence (lower panels in Figures 3-6); and 2) at the time of maximum tidal strain, just before the time of the earthquake (upper panels in Figures 3-6); we call this the "pre-earthquake time". The typical difference between these two times is of a few hours and it is never larger than 12 hours (the time between semidiurnal peaks). Our motivation to include the "pre-eartquake time" is the exploration of the hypothesis that the tidal triggering could not be instantaneous.

It is worth noticing that taking the strain values at the maximum, may or may not contribute to produce larger values of the normal and shear stresses (see Eqs. 3 and 4). Although it is obvious that when the tidal strain $e_{ij}$ values are at their maximum, they are larger than at the time of the Earthquake (with the exception of the case when the earthquake happens exactly at the maximum), previous calculations show that they do not significantly differ from each other (see Moncayo et al. 2019); besides, according to the orientation of the nodal plane, those larger pre-earthquake strains may also correspond to a negative and larger (in absolute value) TCFS. Globally, we notice that in all the studied clusters, the distribution of TCFS is more biased toward positive values when the stresses are calculated at the pre-earthquake maximum, supporting the hypothesis of a non-instantaneous effect of tides on earthquake triggering. These results suggest that the fault zone is influenced by the tides at a time before the earthquake, contributing to nucleate the event.

In the following subsections we present in detail the results for each studied area**.**

**6.1. Bucaramanga**

The histogram and momenta of the TCFS calculated for the Bucaramanga seismic nest are presented in Figures 3 to 7 (upper-left panel). To get a sense of the relative magnitudes of strain associated with BT and OTL, we take an earthquake in June 20, 2004; the total magnitude of the vertical strain at the earthquake origin time is 12.59, and the magnitude of the OTL is 2.17, indicating that the ocean tidal effect represents a relatively small contribution to the total tidal strain.

In general, results indicate that a large friction coefficient is more consistent with the hypothesis of tidal triggering. When the pre-earthquake strains are used to calculate the TCFS, we find a positive mean TCFS value with a dispersion of the same order of magnitude (Figure 3, upper-right panel). In this case, also the skewness and the kurtosis are positive, which is also consistent with a delayed tidal triggering. The box plots in Figure 7 reaffirm these observations. We see that when calculating stresses at pre-earthquake times (upper-left panel, left box), the median of the TCFS values is well above zero. Intriguingly, however, the computed TCFS have atypical values at both sides from zero.

The scatter plots in Figures 8 and 9 show the phase of the vertical strain signal, for the diurnal and monthly phases as plotted against the TCFS for two different friction coefficients. We choose the Bucaramanga nest for this analysis because it has been the most studied one out of the four selected clusters, and because the mechanism of earthquake generation has been most widely discussed (Zarifi et al., 2007, Prieto et al., 2012).

An interesting behavior is evident in Figure 8 and the upper panels of Figure 9. For the diurnal phase, in both cases, phases close to zero or one correspond to mostly positive values of the TCFS. On the other hand, negative TCFS values are only observed for phases around 0.5. These results are consistent with the tidal triggering hypothesis, since as observed and explained in Moncayo et al. (2019), phases close to 0 and 1 correspond to times when tides are especially larger in the tidal cycle. Moreover, in Moncayo et al. (2019) we found that the tidal component that best correlates with the occurrence of earthquakes is the monthly phase; Figures 8 and 9 show that for this phase, the correlation between the TCFS values and the tidal phases are larger for the monthly than for the diurnal component, confirming our expectations.

According to Curchin and Pennington, (1987), the Bucaramanga nest does not show a relationship between the semidiurnal tides and the earthquake origin time; however, they did not include the monthly phase in their calculation, which is the one that most strongly suggests the effect of the tide to facilitate slip along the fault planes defined by the focal mechanisms.

### 6.2. Cauca

Histogram and statistics of the TCFS for the Cauca cluster are presented in Figure 4 and 7 (upper-right panel). A similar picture as that observed in the Bucaramanga seismic nest arises. At pre-earthquake times, large mean values of TCFS with a dispersion of one order of magnitude greater than the average, and a skewed distribution, are observed in this region. At earthquake origin time the TCFS values are negative, however the earthquake happened, this could reinforce the idea about the pre-earthquake forcing. The results in Figure 4 for the Cauca cluster are also shown as box diagrams in the upper-right panel of Figure 7. In this case, the hypothesis of a pre-earthquake forcing, strongly supports the possibility of a tidal contribution to earthquake triggering, while results at the earthquake origin time are inconclusive. The bias of TCFS for events in this area toward positive values, is even more significant than for the case of the Bucaramanga seismic nest, when considering pre-earthquake times. However, at the earthquake origin time, the median of the TCFS values is negative.

### 6.3. El Puyo and Pucallpa

Histogram and statistics of the TCFS for the El Puyo and Pucallpa clusters are presented in Figures 5, 6 and 7 (lower panels).

Considering the TCFS calculated at the time of maximum strain before the earthquake, and using friction coefficients of 0.2 and 0.4 and a rigidity of 75 Gpa (Figures 5 a, b and 6 a, b), we observe consistently positive values of TCFS. In these areas, the average TCFS calculated at pre-earthquake times is almost twice larger than the dispersion (El Puyo) or once larger (Pucallpa), in contrast with the case of the Bucaramanga and Cauca regions. The skewness of the distribution of TCFS values is positive.

For the TCFS calculated at the earthquake origin time, the median and mean of the distribution are above zero, but not as large as for the case of pre-earthquake times. Figure 7 (lower panels) also shows those trends: for pre-earthquake times, all the TCFS are positive for the El Puyo and the Pucallpa clusters; for the TCFS at the earthquake origin time, most of the data are positive, with a similar trend as that for the previous case.

### 6.4. Effects of uncertainties in the fault parameters and the choice of nodal plane

Given the intermediate depth of the analyzed earthquakes and the impossibility to associate the events with regional or local faults, the ambiguity of the nodal planes in the focal mechanism cannot be easily resolved. For this reason, all our previous results of TCFS calculation consider both nodal planes.

Uncertainties in the focal mechanism information can be of the order of 10-45 degrees, and these variations can affect the obtained TCFS. To explore how these uncertainties affect our previous results, we consider random variations within the mentioned interval in the focal mechanism information and carried out the calculations; for this purpose, we perturbed the strike, dip and rake by random angles within the range of the uncertainties and to re-compute the TCFS, first assuming that the random variation is subtracted from the focal mechanism parameters, and in the other case assuming that such variation is added. **Figure 10** shows the result of these numerical experiments. When we compare these results with those of Figure 3a,b. we see only small changes in the distribution of the resulting TCFS. Still, the tendency in all cases is toward positive values of TCFS, suggesting that this trend is not just the result of chance or errors in the estimation of the focal mechanism.

Additionally, to explore the effects of the choice of nodal plane, we carried out a couple of additional tests. First, we randomly selected, for each focal mechanism, a nodal plane 1 (NP1) and a nodal plane 2 (NP2), and determined the statistical distribution of both planes for all available focal mechanisms. For all cases (all seismic clusters), the tendency in the values of the TCFS is preserved regardless of the chosen nodal plane. In Figure 11 we show one example of this tendency for the case of the Bucaramanga nest, where the difference in the mean value for both planes is 36.57 Pa. The second test considers the calculated differences between the TCFS for each nodal plane from the same focal mechanism (See supplementary Figures S1 and S2). The differences in TCFS between the two nodal planes range between 68 to 688 Pa for all the studied clusters when considering the time at the maximum tidal strain before the earthquake, and from 36 to 108 Pa for the case of the TCFS calculated at the earthquake origin time. In the case of the calculations at pre-earthquake time, the Cauca cluster has the smallest differences in the TCFS calculated for the two possible nodal planes per focal mechanism, which are typically less than 100 Pa; El Puyo cluster shows the greatest differences, reaching up to 688 Pa. Despite these differences in the TCFS calculated for both nodal planes, the global trends of the TCFS are similar.

## 7. Discussion

In general, the statistical trends found for the TCFS suggest that there is a contribution of the tides to seismic triggering in intermediate depth earthquake clusters. The focal mechanisms are very diverse, including all types of faulting. This diversity of mechanisms and fault plane orientations may allow that in many cases the gravitational forces from the Moon and the Sun could provide a loosening effect on the fault surfaces, which could facilitate sliding. Our results are consistent with the possibility that this pre-earthquake effect may help triggering seismic events in the clusters. At the earthquake origin time, the general statistical trend of the TCFS also indicates that it contributes to the triggering, although such trend is not as strong as for the case of pre-earthquake times (at the time of maximum tidal strain a few hours before the earthquake).

Two plausible processes that make intermediate depth earthquakes possible are dehydration embrittlement and thermal shear runaway (Frohlich and Nakamura, 2009, Prieto et al., 2012); slip and/or rupture generally occur within cold and stressed subducted lithosphere. Those processes can weaken pre-existing faults, or can even create new planes along which slip can occur, where shear strain localizes. If we also consider the effects of slab concavity, changes of dip along strike, slab pull, mantle flow, possible slab tearing and even collisions between different slabs, the result is a large heterogeneity of focal mechanisms, and therefore, a wide variety of fault planes with diverse attitudes concentrated in a relatively small volume. We hypothesize that this is the reason why the correlation between tides and seismicity is preferentially strong for the case of intermediate depth earthquake clusters, as found by Moncayo et al. (2019). In these clusters, the diversity of fault plane orientations increases the likelihood that the loosening effect of the tides along fault surfaces takes place. For the case of shallower earthquakes, which mostly occur on pre-existing regional or local faults, the greater homogeneity of plane attitudes makes the tides less likely to contribute to earthquake triggering.

Even though previous studies and focal mechanism data for the analyzed seismic clusters in the northern-central Andes are relatively scarce, for the cases of the Bucaramanga nest, and El Puyo and Pucallpa clusters, it seems that there is a tidal contribution to earthquake triggering; for the case of the Cauca cluster, evidence is less conclusive. It is worth noticing that the Cauca cluster is the most spatially disperse of the analyzed seismic accumulations. The other three clusters are clearly located within a subducted slab and confined within a volume smaller than a cube of side length 100 km, with a significantly smaller volume for the Bucaramanaga nest (Prieto et al., 2012, Yepes et al., 2016, Soles-Valdivia, 2012).

Events in Bucaramanga may be associated with slab tearing (Cortes and Angelier, 2005), slab collision (Zarifi et al., 2007), reactivation of pre-existing fault planes or localized accumulation of shear strain within the slab (Prieto et al., 2012). On the other hand, the Cauca cluster occupies a larger volume (Chang et al., 2017, 2019) and has some events in the mantle wedge above the subducting slab. Moreover, they mostly occur along a vertical nodal plane that may allow the fluid propagation that is linked with the earthquake generation. Since there are specific directions along which slip preferentially occurs, it is not surprising that the causal relation between tides and earthquakes is more difficult to establish for this cluster, as the tidal forces have a more limited variety of plane attitudes on which they can generate the loosening effect, and therefore, the likelihood of contributing to earthquake triggering decreases.

## 8. Summary and Conclusions

For the main four accumulations of intermediate depth seismicity in the northern – central Andes, we calculated the Tidal Coulomb Failure Stress (TCFS) for each of the events within the clusters with available source parameter information. The TCFS includes the effects of normal and shear stress on a fault plane, and when positive, it indicates favorability to promote slip along the fault.

We calculated the values of TCFS at the earthquake origin time and at the time of maximum strain a few hours before the earthquake. Although a small number of events with focal mechanism information for the selected seismic clusters were available (56 for Bucaramanga nest, and 89, 23 and 28 for Cauca, El Puyo and Pucallpa clusters respectively), the general results of our calculations of TCFS suggest that tides may contribute to earthquake triggering at seismic clusters of intermediate depth.

Since the Bucaramanga nest is the most studied one and has the clearest connection between tides and seismicity, we performed additional tests to elucidate how such a connection may work. Since the results of TCFS at pre-earthquake times show a clear trend towards positive values, we consider that the possibility of a pre-earthquake tidal forcing that may help trigger the event is plausible, so the fault can be critically affected, and once it reaches a critical stage, smaller levels of tidal strain could bring the fault to slip.

We had previously found that for northwestern South America a statistical correlation between earthquakes and tides is hinted for intermediate depth clusters, but it is not for upper crustal events (Moncayo et al., 2019). In this work, we evidenced that tidal stresses can contribute to the physical processes that promote earthquakes within those clusters. What makes the intermediate depth clusters particular, is the large variability of fault plane orientations within a small volume, which can increase the likelihood that tidal stresses can contribute to earthquake triggering. Upper crustal earthquakes are usually associated to very specific fault plane orientations, so the tidal action is more limited.

Among the four analyzed clusters in this study, the Cauca cluster is the one with the less clear contribution of the TFCS to earthquakes, which is likely due to the fact that this is a relatively disperse cluster with some events located above the subducting slab that preferentially occur along one or several near-vertical planes (Chang et al., 2019).

**Aknowledgements**

G.Moncayo wants to thank the Administrative Department of Science, Technology and Innovation of Colombia COLCIENCIAS, for their financial support during the year 2016. Many thanks to Dr. Thomas Jahr and PhD Student Chris Salomon from the Institute of Geosciences of the Friedrich-Schiller University (FSU) of Jena (Germany), for the useful discussions and support during the 3-month internship of G. Moncayo at this University (September to December 2016); as well as to Dr. Adelheid Weise (LIAG-Leinbniz Germany) and Prof. i.R. Dr. Gerhard Jentzsch (FSU) Jena for the useful recommendations in relation with this work and support with ETERNA 3.40. Additionally, many thanks to Dr. Walter Zuern from Black Forest Observatory (Germany), Dr. Kogi Matsumoto (National Astronomical Observatory-japan, GOTIC2 program) for the personal communications, and to Jorge Castro at the Mariana University in Pasto for the support with some statistical suggestions for the work. Discussions with Dr. Ludger Suarez from the National University of Colombia greatly helped to improve this research.

We thank the Colombian Geological Survey and the National Seismological Network of Colombia (RSNC), for the free available earthquake database and some personal communications. We thank the Global Centroid Catalog too, for the free accessibility to focal mechanism information.

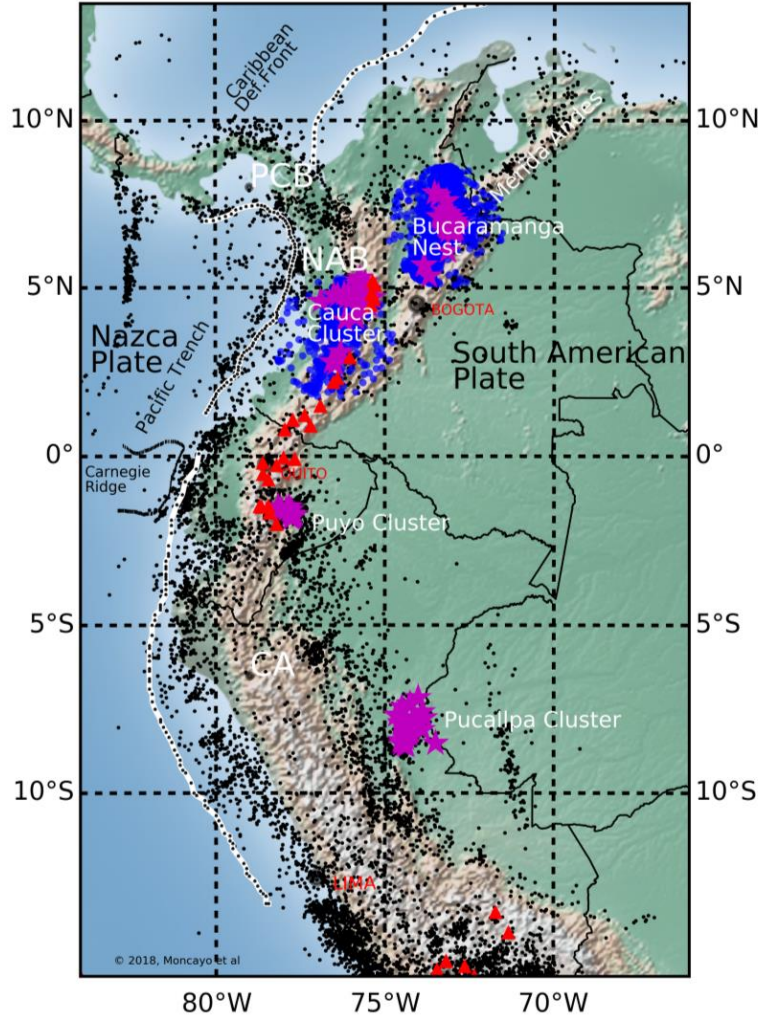

Figure 1. The North Andean Block and the Central Andes (NAB and CA respectively). Black dots show the seismic activity in the region, obtained from the National Earthquake Center (NEIC) and from the Centroid Moment Tensor catalog (CMT), between 1993 and 2018. Blue circles represent all the reported seismic events in the Cauca seismic cluster and in the Bucaramanga seismic nest, considering a radius of 200 km around the center of each nest (lat. 4.5ºN, lon. 76ºW and lat. 7ºN, lon. 73ºW, respectively) and a depth greater than 100 km. Magenta stars correspond to the seismic events with focal mechanism information in the Bucaramanga nest, and the Cauca, El Puyo and Pucallpa clusters. Red triangles represent volcanoes. PCB: Panama-Choco Block, NAB: North Andean Block, CA: Central Andes, Caribbean Def. Front: Caribbean Deformation Front.

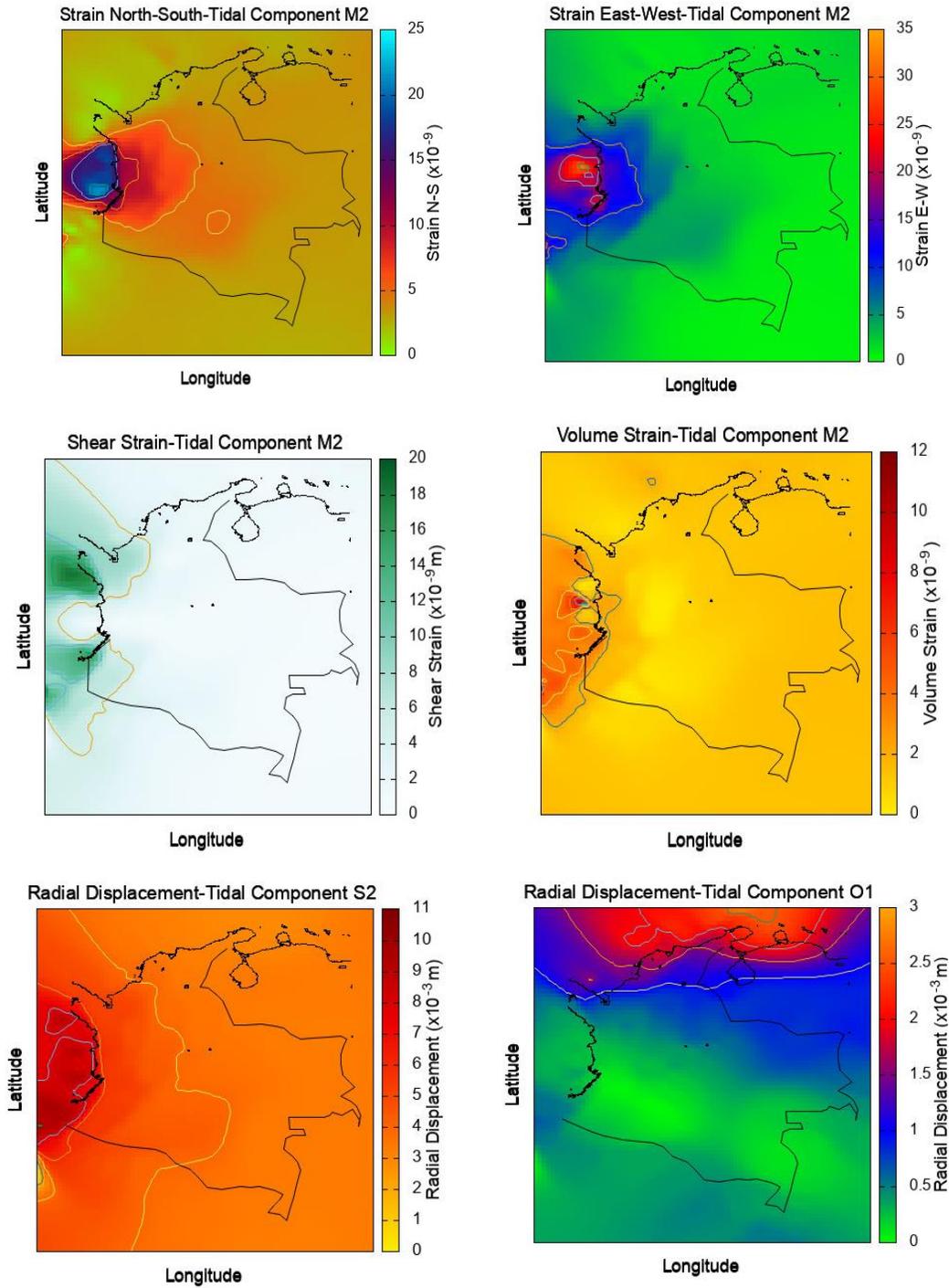

Figure 2. First and second row: OTL strain components (North-South, East-West, shear and volumetric) for Colombian territory and neighboring regions in Northern South America. We plot, for reference, only the strain components for the tidal constituent M2 (principal lunar semidiurnal). Lower row: OTL radial displacement for the short period tidal constituents S2 (principal solar semidiurnal) and O1 (lunar declinational diurnal).

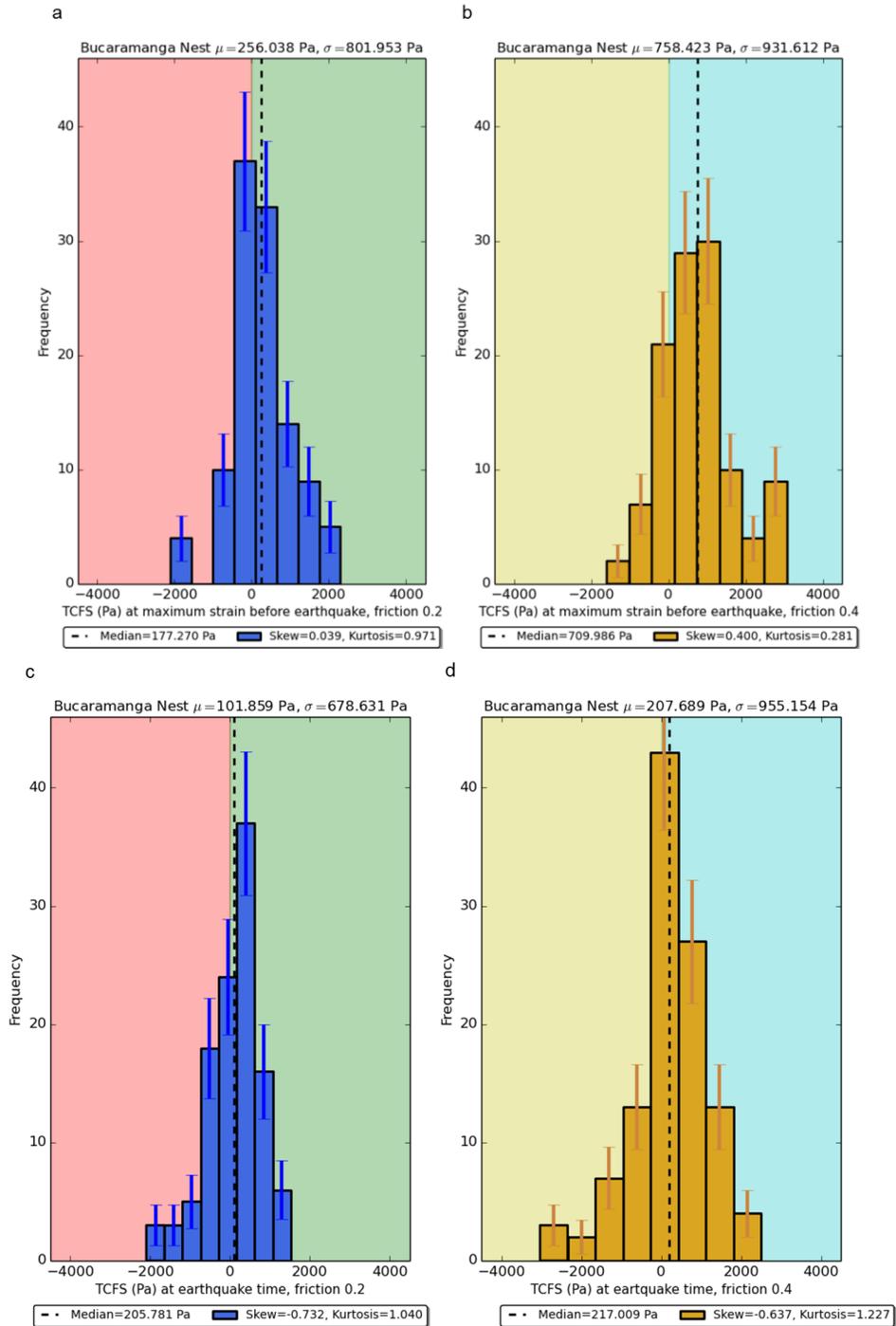

Figure 3. Frequency histograms of the TFCS for the Bucaramanga seismic nest. (a) and (b) show the TCFS calculated at the time of maximum strain before the earthquake occurrence, with friction coefficients of 0.2 and 0.4 respectively. (c) and (d) show the TCFS calculated from tidal strain data at the earthquake origin time, with friction coefficients of 0.2 (c) and 0.4 (d). The histograms include information of the solid Earth tides and OTL on both possible nodal planes per earthquake, and the rigidity is assumed to be 75 GPa. On top of each histogram we report the mean and standard deviation.

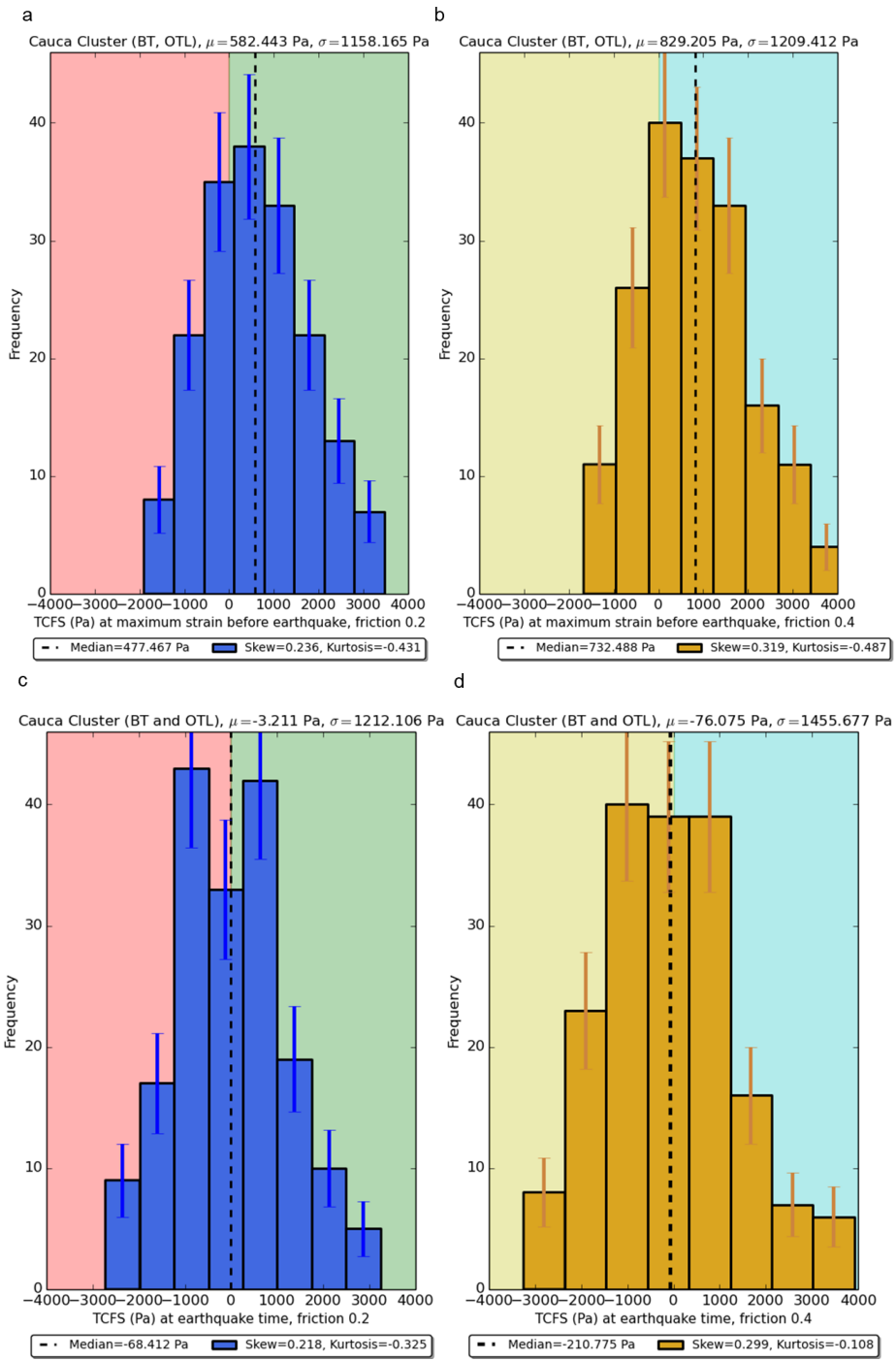

Figure 4. Frequency histograms of the TFCS for the Cauca seismic cluster, considering BT and OTL. Plot structure, conventions and assumptions are the same as in Figure 3.

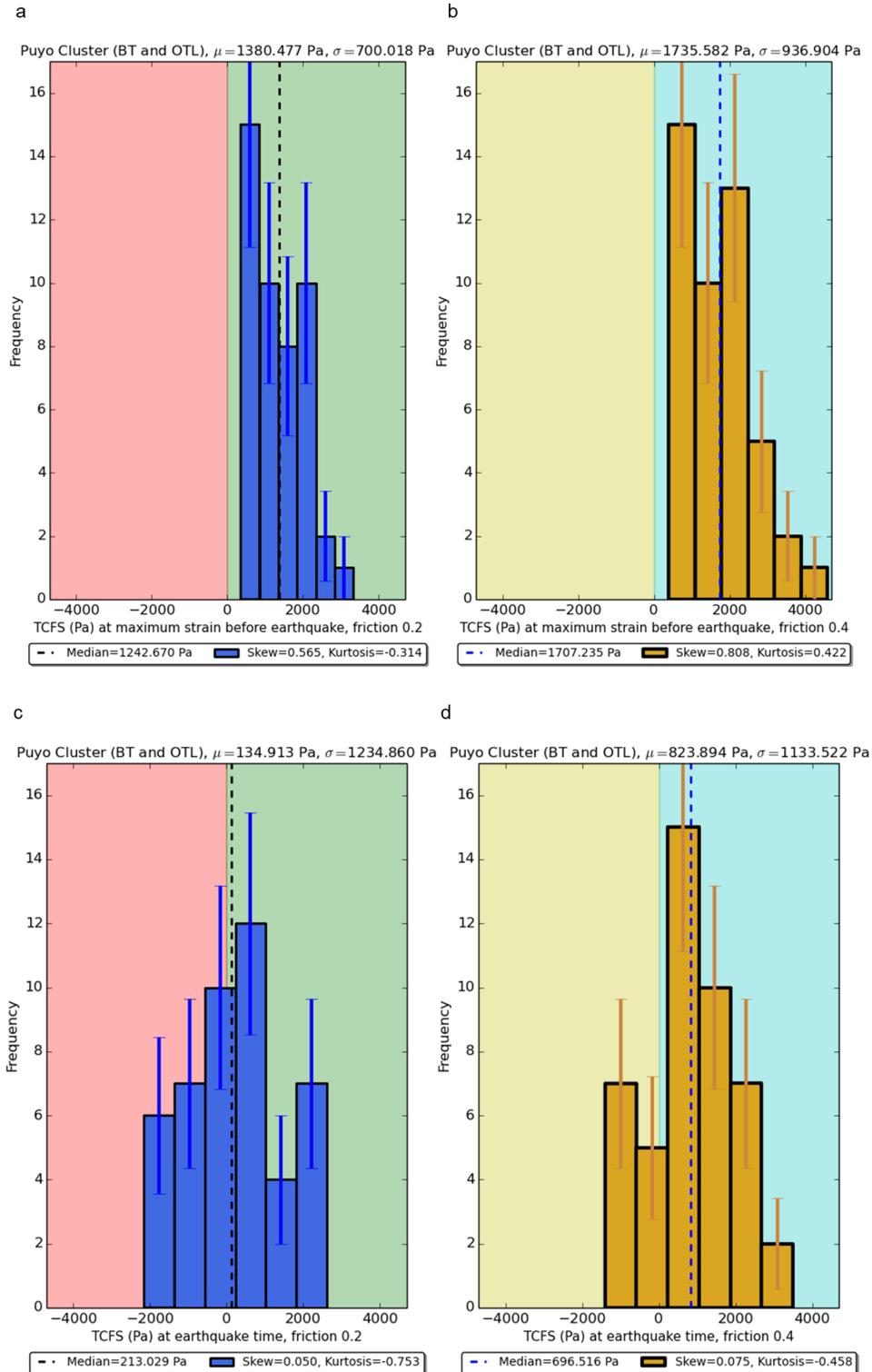

Figure 5. Frequency histograms of the TFCS for the El Puyo (Ecuador) seismic cluster. Plot structure, conventions and assumptions are the same as in Figure 3.

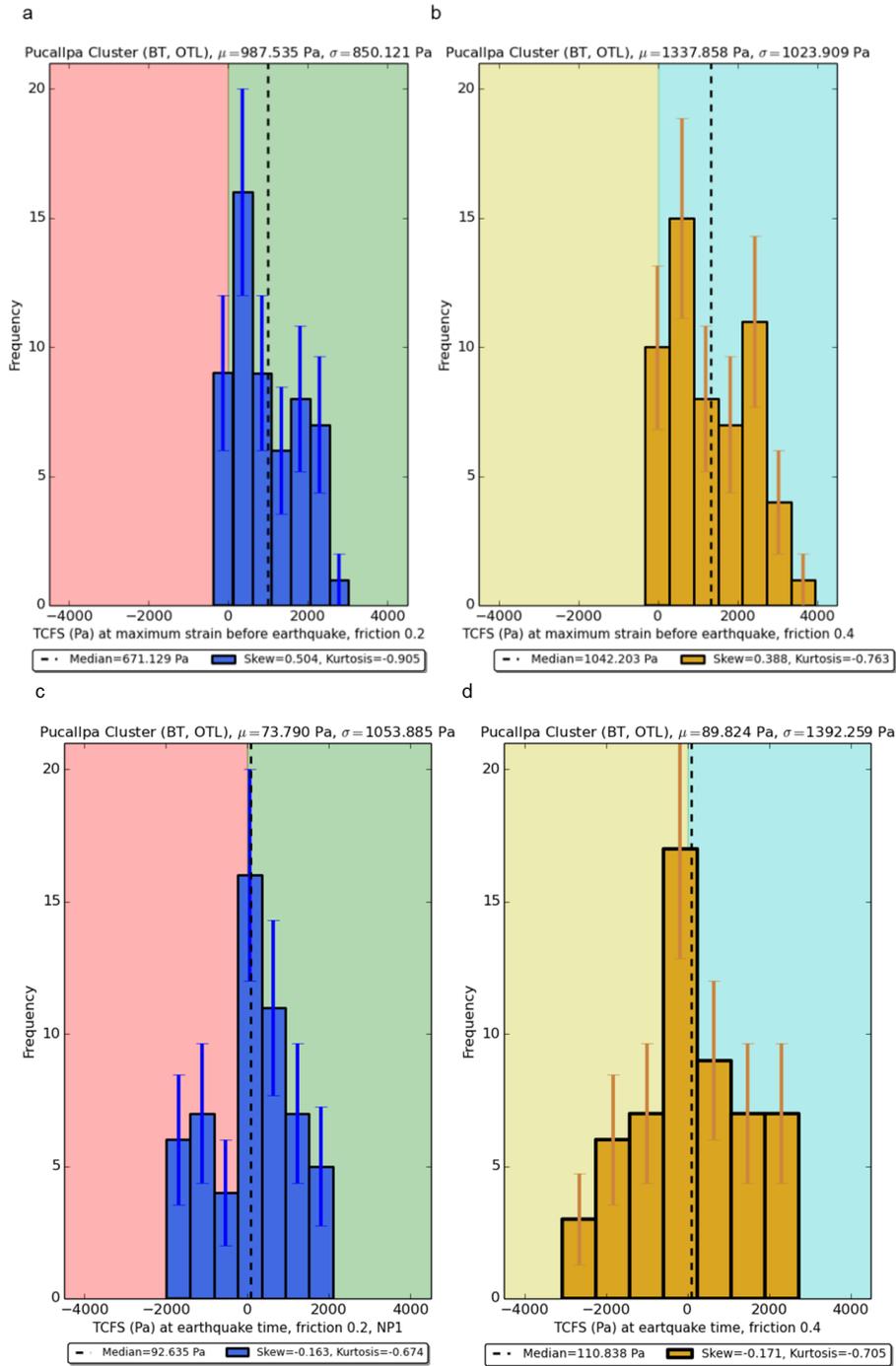

Figure 6. Frequency histograms of the TCFS for the Pucallpa (Peru) seismic cluster. Plot structure, conventions and assumptions are the same as in Figure 3.

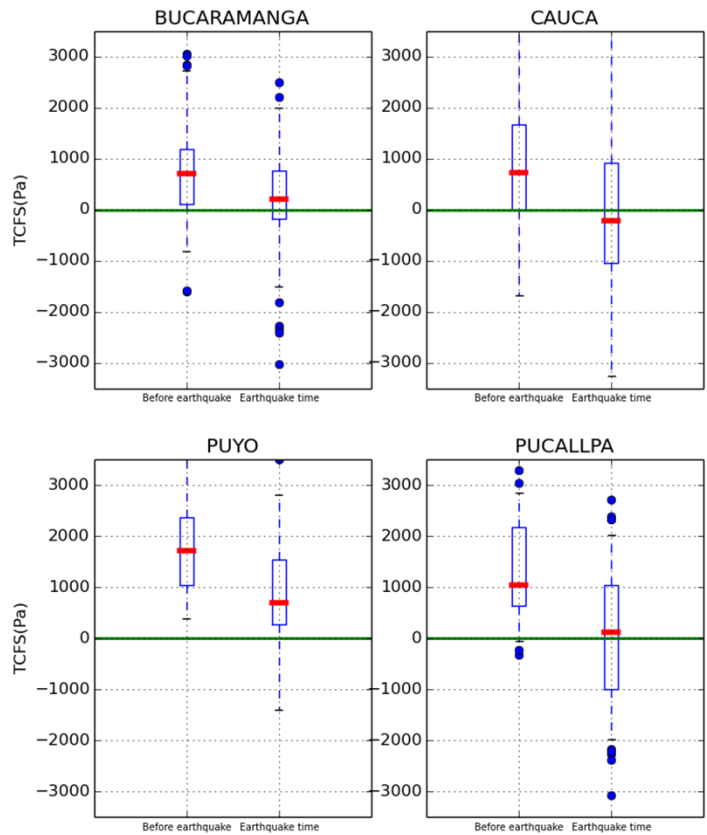

Figure 7. Box plots for the Bucaramanga, Cauca, El Puyo and Pucallpa seismic clusters. Each panel includes information at the time of maximum strain before the earthquake (left) and at the earthquake origin time (right). The box plots correspond to the analysis made with a friction coefficient of 0.4 and for the combined effect of BT and OTL. The orange lines show the median of the sample, which coincides with the second quartile. Inside each box we found the 50% of the sample. The blue circles represent atypical data, which are located at a greater distance than 1.5*IQR (IQR=interquartile range).

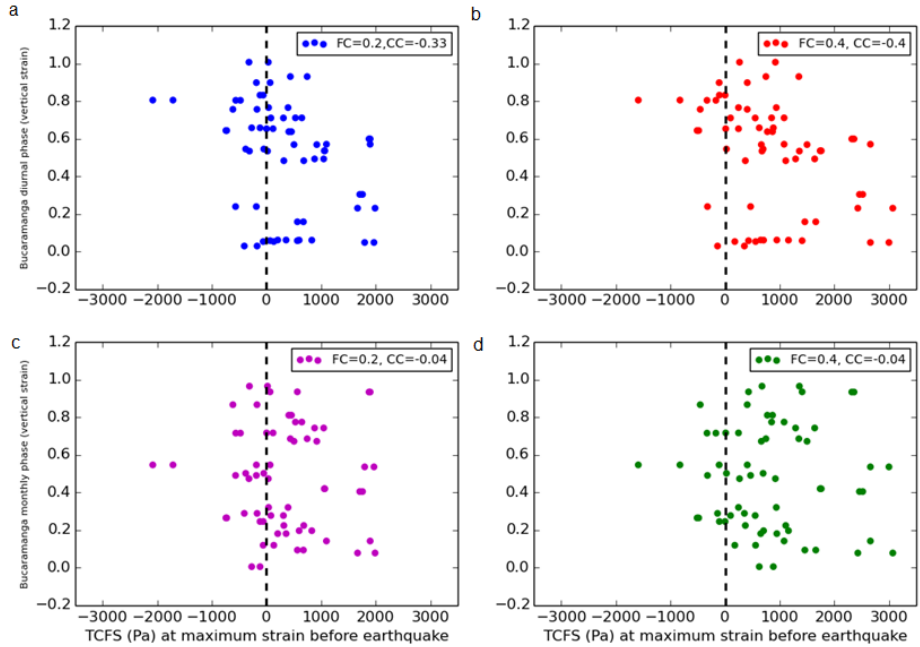

Figure 8. Scatter plots of vertical strain vs the TCFS for the Bucaramanga seismic nest, using friction coefficients of 0.2 and 0.4, and a rigidity of 75 GPA; (a) and (b) consider the diurnal phase, and (c) and (d) the monthly phase. For the TCFS calculations we considered the time of maximum strain before the earthquake origin time. FC: Friction Coefficient; CC: Correlation Coefficient.

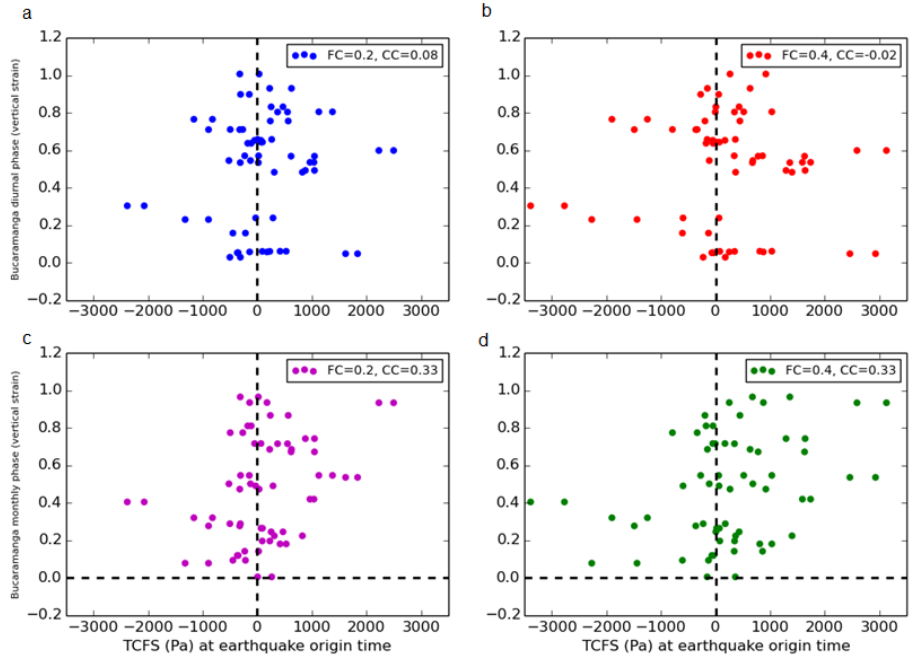

Figure 9. Same as in Figure 8, but using the strain at the earthquake origin time.

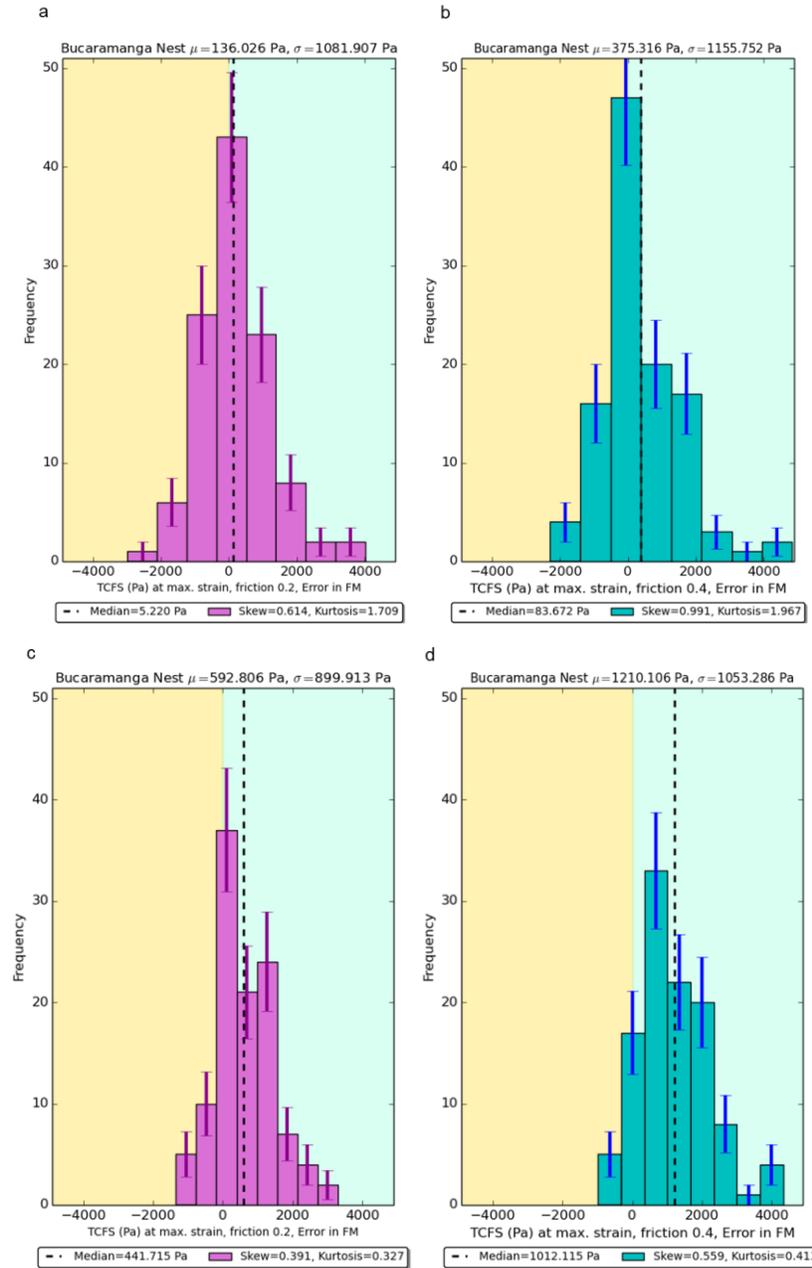

Figure 10. Histograms of the TFCS for the Bucaramanga seismic nest considering both nodal planes and the TCFS calculated when random variations between 10 and 45° in the focal mechanism are introduced. (a) and (b) correspond to the TCFS calculated when the random angle is negative. Panels (c) and (d) show the TCFS calculated when the angle variations are added. Friction coefficient is 0.2 for (a) and (c), and 0.4 for (b) and (d). The TCFS calculations include both BT and OTL. Rigidity is assumed to be 75 GPa. On top of each histogram we report the mean and standard deviation.

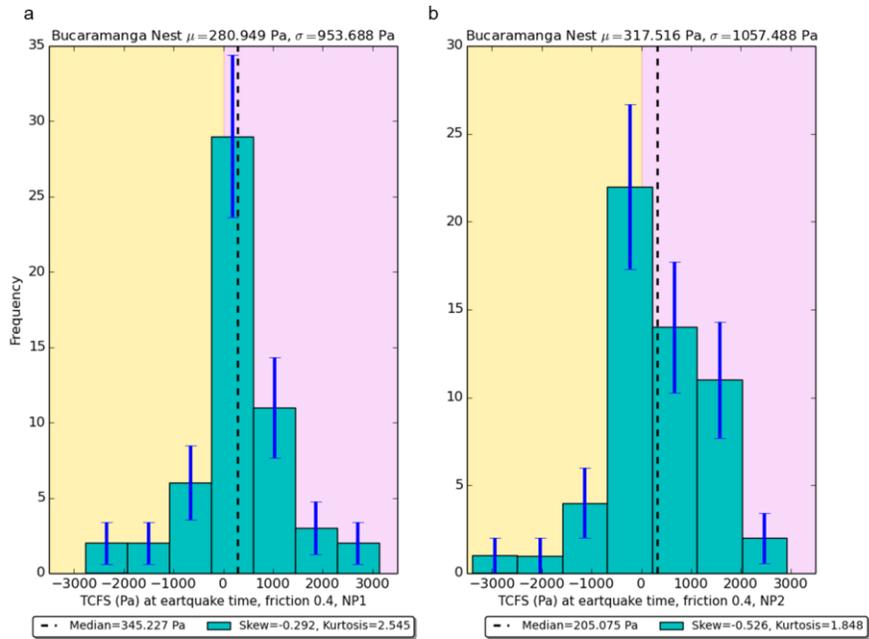

Figure 11. TCFS for randomly determined (a) nodal plane 1 (NP1) and (b) nodal plane 2 (NP2) from each available focal mechanism for the Bucaramanga nest.

Supplementary Material

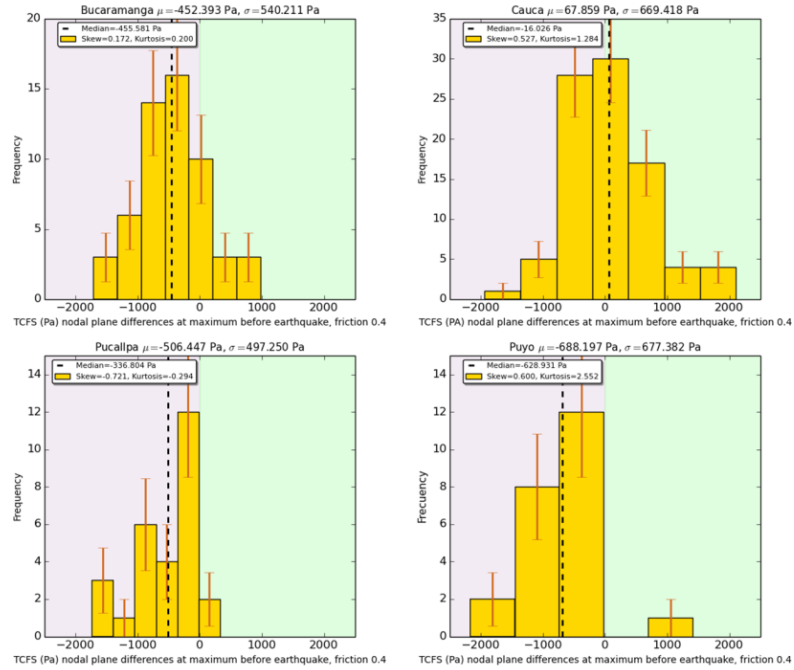

Figure S1. Differences between the nodal planes for the calculated TCFS and for all seismic clusters. The plots correspond to the time at the maximum tidal strain before the earthquake, and a friction coefficient of 0.4.

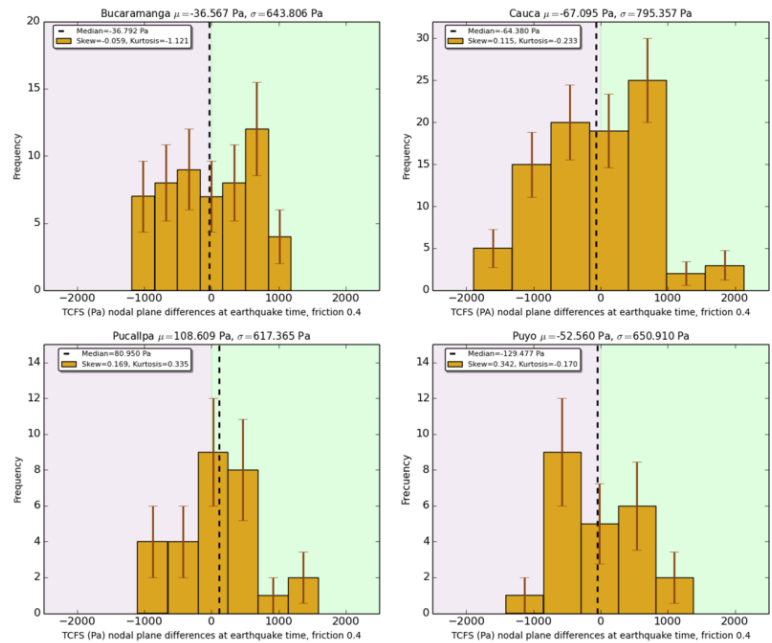

Figure S2. Histograms of differences in the TCFS when calculated with both nodal planes for each focal mechanism. Results are shown for all seismic clusters. These plots correspond to the earthquake time and a friction coefficient of 0.4.

Table S1. Calculated normal $\sigma_N$ and shear $\sigma_S$ tidal stress for 56 intermediate-depth events from 1977 to 2017 in the Bucaramanga seismic nest. At maximum strain before earthquake = MSBE and at earthquake origin time = AEOT. The data 1-39 combines information of Cortes and Angelier (2005) with information of the RSNC and CMT Catalog (Dziewonski et al., 1981; Ekström et al., 2012). NP1 corresponds to the strike (S), dip (D) and rake (R) of the first nodal plane. Date are in Julian Days.

| No. | Julian Day | Lat º | Lon º | Depth (km) | Ml | NP1 | | | AEOT | | MSBE | |
|---|---|---|---|---|---|---|---|---|---|---|---|---|
| | | | | | | S | D | R | $\sigma_N$ (Pa) | $\sigma_S$ (Pa) | $\sigma_N$ (Pa) | $\sigma_S$ (Pa) |
| 1 | 2443438,56 | 7,58 | -73,17 | 169,3 | 5,2 | 196 | 13 | -52 | 167,94 | 540,86 | 187,60 | 574,89 |
| 2 | 2443225,59 | 7,17 | -73,18 | 157,7 | 5,5 | 15 | 46 | 160 | -1217,57 | 658,20 | 1981,54 | -308,09 |
| 3 | 2443888,31 | 6,85 | -73,21 | 157,4 | 5,1 | 261 | 8 | -29 | 32,60 | 59,54 | 82,21 | 289,92 |
| 4 | 2443944,01 | 7,03 | -73,32 | 157,7 | 5,7 | 118 | 40 | 170 | -768,16 | -204,59 | 1895,61 | -719,57 |
| 5 | 2444467,40 | 6,67 | -73,12 | 166,0 | 5,1 | 257 | 15 | -2 | 116,87 | 40,93 | 272,18 | 37,21 |
| 6 | 2445474,47 | 7,11 | -73,41 | 155,1 | 5,2 | 350 | 76 | -14 | 2634,57 | -140,66 | 1721,81 | 420,17 |
| 7 | 2445575,85 | 6,64 | -72,97 | 156,3 | 5,3 | 278 | 34 | 18 | 1285,52 | -577,16 | 1365,00 | -581,29 |
| 8 | 2446009,05 | 7,28 | -73,25 | 155,7 | 5,4 | 187 | 14 | -96 | 149,52 | 614,05 | 187,39 | 746,75 |
| 9 | 2446403,24 | 7,10 | -73,34 | 153,4 | 5,3 | 34 | 59 | 179 | 21,88 | 654,38 | 1917,02 | 202,40 |
| 10 | 2446243,49 | 7,10 | -73,00 | 144,2 | 5,1 | 12 | 9 | 76 | 75,63 | -493,43 | 83,31 | -506,77 |
| 11 | 2446734,80 | 7,21 | -73,13 | 153,2 | 5,2 | 131 | 33 | -171 | -190,18 | 309,38 | 1953,87 | -15,37 |
| 12 | 2446611,34 | 7,15 | -73,14 | 160,1 | 5,7 | 259 | 46 | 18 | 207,93 | 224,63 | 1393,31 | -687,67 |
| 13 | 2447693,16 | 6,87 | -73,09 | 156,8 | 5,2 | 109 | 8 | -173 | 41,33 | 43,40 | 63,26 | 47,46 |
| 14 | 2448228,53 | 6,78 | -72,95 | 158,2 | 5,3 | 315 | 21 | 71 | -383,28 | 914,76 | 417,74 | -925,97 |
| 15 | 2448624,32 | 5,67 | -73,84 | 146,0 | 5,7 | 170 | 55 | 180 | -1251,26 | -644,18 | 922,14 | -11,29 |
| 16 | 2448987,56 | 6,74 | -72,96 | 154,3 | 5,1 | 41 | 42 | 124 | -748,60 | 418,51 | 798,50 | -592,11 |
| 17 | 2449697,14 | 6,89 | -72,85 | 164,4 | 5,2 | 45 | 42 | 121 | -449,10 | 299,20 | 1137,49 | -967,01 |
| 18 | 2449858,43 | 5,66 | -73,84 | 141,1 | 5,2 | 282 | 17 | -7 | 257,92 | 244,58 | 252,26 | 130,95 |
| 19 | 2450449,69 | 6,98 | -72,91 | 164,3 | 5,3 | 126 | 48 | 146 | -947,20 | 603,00 | 1151,32 | -754,77 |
| 20 | 2451490,74 | 6,90 | -73,15 | 160,2 | 5,4 | 54 | 38 | 152 | 1593,44 | -863,30 | 2066,15 | -855,46 |
| 21 | 2452233,38 | 6,74 | -72,90 | 153,7 | 5,3 | 324 | 73 | -9 | 568,37 | 112,34 | 542,14 | 490,63 |
| 22 | 2453281,85 | 6,83 | -73,04 | 172,4 | 4,8 | 148 | 64 | -170 | 1614,61 | 177,39 | 2215,22 | -227,72 |
| 23 | 2453175,13 | 6,82 | -73,01 | 151,2 | 4,8 | 353 | 36 | 143 | 154,76 | -336,14 | 321,59 | -167,94 |
| 24 | 2453426,78 | 6,88 | -73,09 | 159,2 | 5,3 | 57 | 54 | -176 | 3641,45 | 544,03 | 4319,13 | 838,16 |
| 25 | 2453745,63 | 6,89 | -73,20 | 171,6 | 5,0 | 32 | 43 | 151 | 1308,47 | -768,08 | 1355,92 | -704,42 |
| 26 | 2453864,13 | 7,02 | -73,24 | 168,4 | 4,9 | 71 | 66 | 180 | 2995,17 | 337,23 | 3414,71 | 380,92 |
| 27 | 2454226,70 | 6,84 | -73,13 | 156,6 | 5,4 | 36 | 49 | 173 | 493,27 | 676,69 | 1636,59 | 57,85 |
| 28 | 2454514,39 | 6,82 | -73,04 | 156,9 | 5,4 | 10 | 45 | 165 | -2116,13 | -356,70 | 1010,95 | -122,97 |
| 29 | 2455132,18 | 6,93 | -73,07 | 164,2 | 5,1 | 42 | 54 | 175 | -9,93 | -174,72 | 1808,85 | 49,80 |
| 30 | 2454941,87 | 7,81 | -73,53 | 129,6 | 5,1 | 300 | 28 | 40 | -559,03 | 685,95 | 779,25 | -769,16 |
| 31 | 2455982,07 | 6,84 | -73,02 | 150,5 | 5,1 | 226 | 42 | 52 | -1227,84 | 1122,94 | 2464,00 | -2581,42 |
| 32 | 2456171,71 | 6,86 | -73,12 | 164,4 | 5,1 | 48 | 43 | 151 | 2866,96 | -910,28 | 3009,98 | -1017,26 |
| 33 | 2456499,01 | 6,87 | -73,13 | 155,1 | 5,0 | 72 | 67 | -166 | -4794,09 | -362,47 | 5328,48 | 934,76 |
| 34 | 2456932,15 | 6,82 | -73,16 | 143,1 | 4,3 | 118 | 67 | -153 | -1799,18 | -368,95 | 2224,89 | 303,13 |
| 35 | 2456981,74 | 6,83 | -73,17 | 157,0 | 4,3 | 355 | 54 | -60 | 1297,43 | 792,54 | 2744,62 | 1752,75 |
| 36 | 2456987,35 | 6,89 | -73,02 | 168,3 | 5,0 | 146 | 48 | 42 | 1326,00 | -625,92 | 1535,90 | -551,67 |
| 37 | 2457005,58 | 5,34 | -73,69 | 158,5 | 4,8 | 250 | 9 | -18 | -40,58 | -66,60 | 34,09 | 5,82 |
| 38 | 2456696,36 | 6,81 | -73,07 | 159,3 | 5,3 | 13 | 22 | 91 | -146,15 | 359,60 | 100,40 | -245,71 |
| 39 | 2456739,37 | 6,08 | -73,14 | 152,0 | 4,4 | 77 | 89 | -171 | 3094,98 | 307,22 | 3238,66 | -76,93 |
| 40 | 2456745,69 | 6,83 | -73,16 | 147,7 | 4,4 | 42 | 58 | 9 | 4583,34 | -1048,71 | 4874,87 | -1256,25 |
| 41 | 2456746,86 | 6,80 | -73,16 | 142,2 | 4,4 | 352 | 68 | 40 | -767,58 | 481,20 | 4814,34 | -1175,33 |
| 42 | 2456888,75 | 6,83 | -73,16 | 154,4 | 4,2 | 152 | 23 | -157 | -227,10 | -283,27 | 126,93 | 106,10 |
| 43 | 2457047,93 | 6,82 | -73,15 | 149,6 | 4,5 | 163 | 84 | 25 | 2623,87 | -234,53 | 3227,98 | -102,53 |
| 44 | 2457309,82 | 7,73 | -73,42 | 120,0 | 5,1 | 333 | 44 | 2 | 791,31 | -193,24 | 2139,47 | 253,72 |
| 45 | 2457355,24 | 6,82 | -73,15 | 154,0 | 4,7 | 68 | 70 | -157 | 3530,44 | 789,02 | 4703,55 | 938,07 |
| 46 | 2457092,37 | 6,83 | -73,11 | 155,9 | 6,2 | 142 | 29 | -152 | 639,35 | 482,28 | 808,05 | 409,58 |
| 47 | 2457104,44 | 6,98 | -73,07 | 170,3 | 4,9 | 66 | 52 | 166 | -1553,46 | -221,66 | 4359,07 | -388,33 |
| 48 | 2457144,91 | 6,81 | -73,15 | 141,1 | 4,4 | 105 | 30 | 30 | -1224,47 | 652,17 | 1204,60 | -801,06 |
| 49 | 2457204,75 | 6,80 | -73,13 | 145,2 | 4,2 | 84 | 79 | 33 | 1749,96 | -607,41 | 4018,65 | -395,76 |
| 50 | 2457218,48 | 6,79 | -73,09 | 153,3 | 4,5 | 58 | 65 | -156 | -2976,40 | -1158,32 | 3804,68 | 963,75 |
| 51 | 2457429,80 | 6,82 | -73,14 | 154,2 | 4,5 | 177 | 74 | 42 | 4834,23 | -1036,91 | 4818,05 | -1071,27 |
| 52 | 2457446,73 | 6,83 | -73,13 | 150,0 | 4,2 | 105 | 65 | 44 | 1126,37 | 173,55 | 3198,65 | -776,69 |
| 53 | 2457425,00 | 6,82 | -73,12 | 153,0 | 4,2 | 50 | 48 | 20 | -452,24 | 60,82 | 2660,29 | -1119,62 |
| 54 | 2457456,77 | 6,84 | -73,15 | 148,3 | 4,7 | 130 | 36 | 58 | 2090,41 | -2318,08 | 2546,73 | -2606,93 |
| 55 | 2457508,99 | 6,81 | -73,16 | 145,5 | 4,2 | 329 | 75 | -27 | 2012,59 | 511,30 | 1948,02 | 205,16 |
| 56 | 2457653,95 | 6,89 | -73,10 | 157,8 | 4,8 | 51 | 58 | 19 | 1698,26 | -118,81 | 2458,05 | -851,12 |

Table S2. Calculated normal $\sigma_N$ and shear $\sigma_S$ tidal stress for 56 intermediate-depth events from 1977 to 2017 in the Bucaramanga seismic nest. At maximum strain before earthquake = MSBE and at earthquake origin time = AEOT. The data 1-39 combines information of Cortes and Angelier (2005) with information of the RSNC and CMT Catalog (Dziewonski et al., 1981; Ekström et al., 2012). NP2 corresponds to the strike (S), dip (D) and rake (R) of the second nodal plane. Date in days of the Julian Calendar (JC) since the reference time 4713 B.C.

| No. | Date (days in JC) | Lat ° | Lon ° | Depth (km) | Ml | NP2 | | | AEOT | | MSBET | |
|---|---|---|---|---|---|---|---|---|---|---|---|---|
| | | | | | | S | D | R | $\sigma_N$ (Pa) | $\sigma_S$ (Pa) | $\sigma_N$ (Pa) | $\sigma_S$ (Pa) |
| 1 | 2443438,56 | 7,58 | -73,17 | 169,3 | 5,2 | 337,3 | 79,8 | -98,1 | 3346,56 | 540,83 | 3238,92 | 574,88 |
| 2 | 2443225,59 | 7,17 | -73,18 | 157,7 | 5,5 | 119,2 | 75,8 | 45,8 | -2982,83 | 658,08 | 4433,19 | -307,90 |
| 3 | 2443888,31 | 6,85 | -73,21 | 157,4 | 5,1 | 19,8 | 86,1 | -97,0 | 1391,49 | 59,56 | 2806,25 | 290,00 |
| 4 | 2443944,01 | 7,03 | -73,32 | 157,7 | 5,7 | 215,7 | 83,6 | 50,4 | -955,42 | -204,61 | 4376,27 | -719,64 |
| 5 | 2444467,40 | 6,67 | -73,12 | 166,0 | 5,1 | 348,9 | 89,5 | -105,0 | 1723,55 | 41,00 | 2509,33 | 37,30 |
| 6 | 2445474,47 | 7,11 | -73,41 | 155,1 | 5,2 | 83,5 | 76,4 | -165,6 | 3361,62 | -140,59 | 2097,83 | 420,20 |
| 7 | 2445575,85 | 6,64 | -72,97 | 156,3 | 5,3 | 172,9 | 80,1 | 122,7 | 3040,74 | -577,17 | 3014,93 | -581,32 |
| 8 | 2446009,05 | 7,28 | -73,25 | 155,7 | 5,4 | 13,2 | 76,1 | -88,5 | 2559,98 | 614,01 | 3022,29 | 746,67 |
| 9 | 2446403,24 | 7,10 | -73,34 | 153,4 | 5,3 | 124,5 | 89,1 | 31,0 | -1045,86 | 654,35 | 2352,09 | 202,33 |
| 10 | 2446243,49 | 7,10 | -73,00 | 144,2 | 5,1 | 206,2 | 81,3 | 92,2 | 3430,52 | -493,48 | 3342,26 | -506,76 |
| 11 | 2446734,80 | 7,21 | -73,13 | 153,2 | 5,2 | 33,4 | 85,1 | -57,3 | -737,96 | 309,39 | 6178,66 | -15,19 |
| 12 | 2446611,34 | 7,15 | -73,14 | 160,1 | 5,7 | 156,3 | 77,2 | 134,6 | -841,61 | 224,59 | 2497,67 | -687,58 |
| 13 | 2447693,16 | 6,87 | -73,09 | 156,8 | 5,2 | 12,1 | 89,0 | -82,1 | 2488,75 | 43,32 | 2846,81 | 47,37 |
| 14 | 2448228,53 | 6,78 | -72,95 | 158,2 | 5,3 | 155,3 | 70,2 | 97,1 | -2745,98 | 914,96 | 2447,71 | -926,04 |
| 15 | 2448624,32 | 5,67 | -73,84 | 146,0 | 5,7 | 260,0 | 90,0 | 35,0 | -4417,94 | -644,18 | 2478,14 | -11,29 |
| 16 | 2448987,56 | 6,74 | -72,96 | 154,3 | 5,1 | 178,8 | 56,3 | 63,3 | -431,15 | 418,49 | 1165,51 | -592,03 |
| 17 | 2449697,14 | 6,89 | -72,85 | 164,4 | 5,2 | 186,0 | 55,0 | 65,1 | -388,52 | 299,18 | 1207,16 | -966,99 |
| 18 | 2449858,43 | 5,66 | -73,84 | 141,1 | 5,2 | 18,7 | 88,0 | -106,9 | 2974,55 | 244,48 | 1994,02 | 130,90 |
| 19 | 2450449,69 | 6,98 | -72,91 | 164,3 | 5,3 | 240,3 | 65,5 | 47,4 | -1524,68 | 602,91 | 1465,81 | -754,69 |
| 20 | 2451490,74 | 6,90 | -73,15 | 160,2 | 5,4 | 166,7 | 73,2 | 55,4 | 3503,91 | -863,27 | 3935,46 | -855,40 |
| 21 | 2452233,38 | 6,74 | -72,90 | 153,7 | 5,3 | 56,7 | 81,4 | -162,8 | -102,08 | 112,33 | 1620,06 | 490,63 |
| 22 | 2453281,85 | 6,83 | -73,04 | 172,4 | 4,8 | 53,6 | 81,0 | -26,4 | 1973,63 | 177,41 | 2918,67 | -227,66 |
| 23 | 2453175,13 | 6,82 | -73,01 | 151,2 | 4,8 | 114,4 | 69,3 | 59,9 | 1001,29 | -336,19 | 1590,76 | -168,07 |
| 24 | 2453426,78 | 6,88 | -73,09 | 159,2 | 5,3 | 324,7 | 86,8 | -36,1 | 4894,64 | 544,24 | 5457,51 | 838,35 |
| 25 | 2453745,63 | 6,89 | -73,20 | 171,6 | 5,0 | 144,1 | 70,7 | 50,8 | 2284,83 | -768,16 | 2629,32 | -704,50 |
| 26 | 2453864,13 | 7,02 | -73,24 | 168,4 | 4,9 | 161,0 | 90,0 | 24,0 | 3421,93 | 337,23 | 3439,48 | 380,92 |
| 27 | 2454226,70 | 6,84 | -73,13 | 156,6 | 5,4 | 130,6 | 84,7 | 41,2 | -724,89 | 676,64 | 2981,08 | 57,79 |
| 28 | 2454514,39 | 6,82 | -73,04 | 156,9 | 5,4 | 110,7 | 79,5 | 46,0 | -3639,82 | -356,54 | 2653,54 | -123,08 |
| 29 | 2455132,18 | 6,93 | -73,07 | 164,2 | 5,1 | 134,9 | 86,0 | 36,1 | -470,11 | -174,73 | 2045,83 | 49,92 |
| 30 | 2454941,87 | 7,81 | -73,53 | 129,6 | 5,1 | 173,5 | 72,4 | 112,2 | -2079,96 | 685,83 | 2864,88 | -768,99 |
| 31 | 2455982,07 | 6,84 | -73,02 | 150,5 | 5,1 | 92,4 | 58,2 | 119,0 | -2076,39 | 1122,91 | 4422,13 | -2581,34 |
| 32 | 2456171,71 | 6,86 | -73,12 | 164,4 | 5,1 | 160,1 | 70,7 | 50,8 | 4405,73 | -910,46 | 4742,28 | -1017,44 |
| 33 | 2456499,01 | 6,87 | -73,13 | 155,1 | 5,0 | 336,4 | 77,1 | -23,6 | -2866,04 | -362,37 | 4048,39 | 934,72 |
| 34 | 2456932,15 | 6,82 | -73,16 | 143,1 | 4,3 | 16,7 | 65,3 | -25,5 | -703,57 | -368,74 | 1917,46 | 303,06 |
| 35 | 2456981,74 | 6,83 | -73,17 | 157,0 | 4,3 | 130,5 | 45,5 | -124,5 | 793,96 | 792,50 | 2440,28 | 1752,66 |
| 36 | 2456987,35 | 6,89 | -73,02 | 168,3 | 5,0 | 24,9 | 60,2 | 129,5 | 1188,72 | -625,97 | 2190,67 | -551,79 |
| 37 | 2457005,58 | 5,34 | -73,69 | 158,5 | 4,8 | 357,8 | 87,2 | -98,6 | -66,01 | -66,61 | 1058,29 | 5,78 |
| 38 | 2456696,36 | 6,81 | -73,07 | 159,3 | 5,3 | 191,9 | 68,0 | 89,6 | -884,78 | 359,62 | 601,43 | -245,70 |
| 39 | 2456739,37 | 6,08 | -73,14 | 152,0 | 4,4 | 346,8 | 81,0 | -1,0 | 2445,70 | 307,22 | 1443,83 | -76,89 |
| 40 | 2456745,69 | 6,83 | -73,16 | 147,7 | 4,4 | 307,2 | 82,4 | 147,7 | 6424,66 | -1048,50 | 6678,10 | -1256,04 |
| 41 | 2456746,86 | 6,80 | -73,16 | 142,2 | 4,4 | 244,6 | 53,4 | 152,2 | -1939,98 | 481,21 | 4810,41 | -1175,38 |
| 42 | 2456888,75 | 6,83 | -73,16 | 154,4 | 4,2 | 40,7 | 81,2 | -68,7 | -3006,52 | -283,23 | 1065,35 | 106,08 |
| 43 | 2457047,93 | 6,82 | -73,15 | 149,6 | 4,5 | 70,2 | 65,2 | 173,4 | 2263,06 | -234,61 | 4254,90 | -102,69 |
| 44 | 2457309,82 | 7,73 | -73,42 | 120,0 | 5,1 | 241,6 | 88,6 | 134,0 | 1007,87 | -193,25 | 5621,31 | 253,65 |
| 45 | 2457355,24 | 6,82 | -73,15 | 154,0 | 4,7 | 329,7 | 68,5 | -21,6 | 1579,85 | 788,98 | 3649,56 | 937,97 |
| 46 | 2457092,37 | 6,83 | -73,11 | 155,9 | 6,2 | 27,1 | 76,8 | -63,9 | 2375,43 | 482,41 | 3000,28 | 409,74 |
| 47 | 2457104,44 | 6,98 | -73,07 | 170,3 | 4,9 | 74,7 | 79,0 | 38,8 | -552,52 | -221,59 | 5154,01 | -388,38 |
| 48 | 2457144,91 | 6,81 | -73,15 | 141,1 | 4,4 | 348,4 | 75,5 | 116,6 | -3373,19 | 652,30 | 3251,15 | -801,22 |
| 49 | 2457204,75 | 6,80 | -73,13 | 145,2 | 4,2 | 346,9 | 57,7 | 167,0 | 1405,38 | -607,43 | 1931,06 | -395,70 |
| 50 | 2457218,48 | 6,79 | -73,09 | 153,3 | 4,5 | 317,3 | 68,4 | -27,0 | -4690,56 | -1158,15 | 3668,12 | 963,72 |
| 51 | 2457429,80 | 6,82 | -73,14 | 154,2 | 4,5 | 73,1 | 50,0 | 158,9 | 4013,47 | -1036,90 | 4266,04 | -1071,25 |
| 52 | 2457446,73 | 6,83 | -73,13 | 150,0 | 4,2 | 352,8 | 51,0 | 147,1 | 143,01 | 173,58 | 1523,15 | -776,62 |
| 53 | 2457425,00 | 6,82 | -73,12 | 153,0 | 4,2 | 306,3 | 75,3 | 136,2 | -2645,80 | 60,67 | 4438,09 | -1119,44 |
| 54 | 2457456,77 | 6,84 | -73,15 | 148,3 | 4,7 | 347,7 | 60,1 | 111,1 | 4180,42 | -2318,11 | 4701,18 | -2606,99 |
| 55 | 2457508,99 | 6,81 | -73,16 | 145,5 | 4,2 | 66,5 | 64,0 | -163,3 | 1687,41 | 511,33 | 2107,71 | 205,23 |
| 56 | 2457653,95 | 6,89 | -73,10 | 157,8 | 4,8 | 310,7 | 74,0 | 146,5 | 2070,02 | -118,86 | 3091,68 | -851,18 |

Table S3. Focal mechanism information for 89 intermediate-depth events from 1966 to 2015 in the Cauca seismic cluster. At maximum strain before earthquake = MSBE and at earthquake origin time = AEOT. The data combines information of Cortes and Angelier (2005) with information from the RSNC, CMT Catalog (Dziewonski et al., 1981; Ekström et al., 2012), Chang et al., 2019, Salcedo et al., 2001, and Tabares et al., 1999. NP1 corresponds to the strike (S), dip (D) and rake (R) of the first nodal plane. Date in days of the Julian calendar (JC) since the reference time 4713 B.C.

| No. | Date (days in JC) | Lat ° | Lon ° | Depth (km) | Ml | NP1 | | | AEOT | | MSBE | |
|---|---|---|---|---|---|---|---|---|---|---|---|---|
| | | | | | | S | D | R | $\sigma_N$ (Pa) | $\sigma_S$ (Pa) | $\sigma_N$ (Pa) | $\sigma_S$ (Pa) |
| 1 | 2439129,26 | 4,65 | -76,00 | 98,0 | 5,1 | 351 | 78 | -157 | -41,41 | -436,89 | 2018,55 | 228,96 |
| 2 | 2441797,28 | 5,20 | -75,80 | 117,0 | 5,3 | 23 | 82 | 21 | -1588,90 | -265,02 | -42,07 | -649,43 |
| 3 | 2441776,08 | 4,70 | -75,80 | 158,0 | 6,1 | 21 | 72 | -32 | -1207,11 | -1184,64 | 4751,82 | -237,83 |
| 4 | 2442515,58 | 4,85 | -75,71 | 139,0 | 5,1 | 232 | 53 | 163 | -906,38 | 915,64 | 3373,42 | 50,10 |
| 5 | 2442918,20 | 4,50 | -75,80 | 166,0 | 5,8 | 40 | 74 | -130 | -2518,00 | -342,96 | 1300,90 | 831,95 |
| 6 | 2444201,49 | 4,81 | -76,20 | 105,0 | 6,3 | 210 | 90 | -150 | -1222,48 | -2275,06 | 1654,28 | 1217,50 |
| 7 | 2444023,04 | 5,28 | -75,73 | 122,0 | 4,9 | 174 | 44 | -100 | -1136,44 | -1250,00 | 144,73 | 146,18 |
| 8 | 2444416,00 | 4,70 | -75,35 | 151,0 | 6,3 | 231 | 74 | 14 | -1310,99 | -757,95 | 2878,29 | -847,81 |
| 9 | 2446518,91 | 4,59 | -75,63 | 163,0 | 5,0 | 230 | 74 | 154 | 722,54 | -1313,41 | 4521,48 | 911,67 |

| No. | Date (days in JC) | Lat º | Lon º | Depth (km) | Ml | NP1 | | | AEOT | | MSBE | |
|---|---|---|---|---|---|---|---|---|---|---|---|---|
| | | | | | | S | D | R | $\sigma_N$ (Pa) | $\sigma_S$ (Pa) | $\sigma_N$ (Pa) | $\sigma_S$ (Pa) |
| 10 | 2447527,37 | 5,10 | -75,78 | 118,0 | 4,9 | 46 | 48 | 27 | -86,81 | -684,81 | 1175,72 | -833,98 |
| 11 | 2448219,44 | 4,45 | -75,46 | 147,0 | 6,0 | 182 | 23 | -139 | 229,15 | 232,35 | 93,46 | 135,68 |
| 12 | 2448850,29 | 5,15 | -75,58 | 107,0 | 5,6 | 288 | 22 | -71 | 664,55 | 1605,21 | 796,55 | 1985,00 |
| 13 | 2450337,77 | 4,59 | -76,90 | 118,2 | 5,3 | 306 | 51 | 144 | 1328,20 | 510,10 | 2732,21 | -2114,97 |
| 14 | 2450794,06 | 4,11 | -75,84 | 189,5 | 6,3 | 249 | 53 | 31 | 2440,73 | -1610,30 | 3405,61 | -1502,15 |
| 15 | 2450499,27 | 4,78 | -76,50 | 118,1 | 5,8 | 138 | 42 | 105 | 884,41 | -701,70 | 1418,54 | -1654,69 |
| 16 | 2452174,64 | 4,03 | -76,17 | 178,0 | 5,9 | 247 | 53 | 31 | -1515,90 | -135,38 | 2799,95 | -2036,72 |
| 17 | 2454722,90 | 4,94 | -75,48 | 139,8 | 5,7 | 227 | 13 | -115 | -242,39 | -1033,93 | 256,19 | 1099,33 |
| 18 | 2455184,10 | 3,08 | -76,30 | 169,2 | 5,1 | 110 | 46 | -172 | -927,25 | -657,99 | 2015,03 | 42,90 |
| 19 | 2455208,28 | 4,29 | -75,76 | 171,4 | 4,7 | 122 | 82 | 125 | -1700,77 | 1070,89 | 1522,97 | -464,78 |
| 20 | 2455226,24 | 4,74 | -76,08 | 127,4 | 4,7 | 13 | 53 | 246 | 3256,54 | 2139,91 | 2958,82 | 2221,28 |
| 21 | 2455199,78 | 3,87 | -76,06 | 112,9 | 3,7 | 210 | 57 | 90 | 3529,60 | -2292,15 | 2888,45 | -1875,78 |
| 22 | 2455247,69 | 4,73 | -76,14 | 99,8 | 3,1 | 45 | 75 | 136 | -677,02 | 899,63 | 2402,94 | 507,00 |
| 23 | 2455269,95 | 4,66 | -76,16 | 93,8 | 2,9 | 104 | 38 | 133 | -1743,31 | 1701,53 | 2321,67 | -2438,98 |
| 24 | 2455258,82 | 4,81 | -76,22 | 110,7 | 4,0 | 10 | 38 | 261 | 2017,26 | 2563,07 | 1176,82 | 1601,64 |
| 25 | 2455300,20 | 4,61 | -76,45 | 97,5 | 3,9 | -51 | 56 | 132 | 3857,05 | -2417,36 | 3841,47 | -2714,51 |
| 26 | 2455395,14 | 4,73 | -76,29 | 107,0 | 2,8 | 190 | 75 | 150 | -876,97 | -108,14 | 2377,71 | -215,12 |
| 27 | 2455464,80 | 4,75 | -76,20 | 95,9 | 3,0 | 29 | 70 | 235 | 3842,80 | 940,65 | 3250,12 | 1842,97 |
| 28 | 2455572,80 | 4,19 | -76,46 | 113,9 | 3,9 | 236 | 75 | 141 | 1118,40 | 1168,18 | 1824,16 | 886,89 |
| 29 | 2455563,72 | 4,52 | -76,27 | 96,8 | 3,3 | 17 | 75 | 158 | 4325,05 | -785,81 | 3635,67 | -299,93 |
| 30 | 2455590,50 | 4,85 | -76,08 | 115,3 | 4,3 | 125 | 75 | 192 | 1263,75 | -231,54 | 970,95 | 90,83 |
| 31 | 2455592,80 | 4,03 | -76,49 | 70,6 | 4,6 | 192 | 51 | 263 | 1085,45 | 875,93 | 2427,74 | 1969,60 |
| 32 | 2455865,76 | 4,79 | -76,14 | 115,0 | 3,5 | 38 | 86 | 248 | 175,64 | 789,00 | 2022,30 | 649,71 |
| 33 | 2455842,20 | 3,87 | -76,10 | 156,2 | 4,2 | 106 | 64 | 115 | 1351,62 | -578,17 | 3428,98 | -1685,97 |
| 34 | 2455876,19 | 4,73 | -76,34 | 101,6 | 3,7 | 138 | 17 | 211 | 238,21 | 11,68 | 338,96 | 340,14 |
| 35 | 2455879,60 | 4,74 | -76,16 | 82,8 | 2,9 | 93 | 36 | 146 | -836,83 | 1055,36 | 1289,11 | -1210,00 |
| 36 | 2455889,47 | 5,10 | -76,40 | 78,0 | 3,9 | 153 | 74 | 148 | -4133,30 | -536,14 | 3341,66 | -1152,65 |
| 37 | 2455898,10 | 4,84 | -76,30 | 90,9 | 2,8 | 126 | 60 | 171 | -89,56 | 1146,97 | 1247,57 | -437,17 |
| 38 | 2455925,36 | 4,79 | -76,35 | 93,4 | 3,1 | 133 | 53 | 219 | 1867,45 | 311,58 | 1712,23 | -138,52 |
| 39 | 2455927,24 | 3,87 | -76,21 | 111,6 | 3,0 | 233 | 68 | 123 | -1712,90 | 1186,68 | 1940,14 | -213,89 |
| 40 | 2455641,07 | 4,90 | -76,22 | 116,2 | 4,4 | 327 | 69 | 204 | -2650,82 | -2192,70 | 5585,21 | -328,93 |
| 41 | 2455672,79 | 4,14 | -76,23 | 140,5 | 3,8 | -108 | 63 | 144 | 3173,74 | -308,36 | 4012,58 | -385,96 |
| 42 | 2455716,41 | 3,97 | -76,72 | 114,5 | 3,4 | 93 | 35 | 162 | -1404,58 | 164,80 | 1487,48 | -692,77 |
| 43 | 2455804,64 | 4,63 | -76,26 | 91,8 | 3,6 | 196 | 75 | 98 | -2085,10 | 788,20 | 5301,20 | -1273,16 |
| 44 | 2455824,16 | 4,31 | -76,33 | 118,4 | 3,5 | 270 | 46 | -180 | -1038,60 | -492,60 | 991,20 | 250,03 |
| 45 | 2455831,48 | 4,71 | -76,19 | 83,3 | 3,0 | 61 | 47 | 165 | -1953,95 | 1028,79 | 4377,96 | -349,52 |
| 46 | 2455958,19 | 4,73 | -76,22 | 76,6 | 2,9 | 282 | 67 | 249 | -1738,92 | -459,71 | 613,05 | -4,31 |
| 47 | 2455931,00 | 4,01 | -76,72 | 100,7 | 3,1 | 102 | 18 | 540 | 45,56 | -338,81 | 147,00 | 63,56 |
| 48 | 2456222,48 | 4,48 | -75,93 | 139,4 | 3,8 | 318 | 79 | 211 | 2359,29 | 906,49 | 2001,00 | 171,63 |
| 49 | 2456207,51 | 4,78 | -76,26 | 110,2 | 4,5 | 154 | 50 | 201 | 580,83 | 856,67 | -349,65 | -706,52 |
| 50 | 2455977,49 | 4,74 | -76,18 | 88,7 | 3,6 | 107 | 33 | 167 | -595,82 | 207,84 | 1502,41 | -805,08 |
| 51 | 2455981,93 | 3,88 | -76,34 | 107,6 | 3,3 | 184 | 65 | 214 | 1530,09 | 465,40 | 1917,47 | 638,80 |
| 52 | 2455987,57 | 4,37 | -76,53 | 96,9 | 4,5 | 140 | 78 | 171 | 265,26 | 1811,46 | 896,01 | -158,21 |
| 53 | 2456009,37 | 4,71 | -76,20 | 86,1 | 2,5 | 126 | 27 | 132 | -326,69 | 1246,74 | 1172,60 | -2143,46 |
| 54 | 2456037,69 | 4,75 | -76,22 | 92,0 | 2,7 | 161 | 73 | 202 | 2904,99 | -250,97 | 2301,58 | -505,62 |
| 55 | 2456088,62 | 4,10 | -76,64 | 68,6 | 3,2 | 103 | 33 | 540 | -961,95 | 69,25 | 1033,56 | -408,27 |
| 56 | 2456105,66 | 3,76 | -76,30 | 143,6 | 4,5 | 136 | 41 | 163 | -680,21 | 725,64 | 1095,23 | -720,96 |
| 57 | 2456149,61 | 3,88 | -76,13 | 153,7 | 3,0 | 256 | 76 | 223 | -2108,14 | -769,97 | 1082,78 | 303,44 |
| 58 | 2456147,23 | 4,05 | -76,15 | 145,3 | 2,8 | 227 | 75 | -180 | -358,24 | 1894,42 | 2049,32 | 1558,45 |
| 59 | 2456148,67 | 4,75 | -76,17 | 91,9 | 2,5 | 269 | 69 | 238 | -2134,28 | -468,50 | 1927,05 | 427,84 |
| 60 | 2456306,83 | 4,97 | -76,02 | 113,8 | 3,0 | 125 | 46 | 224 | 2418,52 | 1818,48 | 2904,68 | 1155,50 |
| 61 | 2456308,58 | 4,48 | -76,20 | 130,4 | 3,0 | 229 | 84 | 171 | -3474,51 | -1404,90 | 3843,95 | 1734,26 |
| 62 | 2456295,20 | 4,15 | -76,40 | 129,2 | 4,4 | -37 | 84 | 124 | -1462,60 | -1057,05 | 2868,50 | -1068,17 |
| 63 | 2456584,52 | 4,90 | -75,90 | 118,3 | 3,0 | 172 | 88 | 207 | -3956,57 | -663,15 | 3701,65 | -118,87 |
| 64 | 2456573,60 | 4,23 | -76,51 | 60,8 | 3,6 | 38 | 54 | 99 | -2809,03 | 2145,05 | 2233,31 | -1298,03 |
| 65 | 2456614,14 | 4,37 | -75,92 | 156,7 | 3,7 | 248 | 71 | 204 | 3318,58 | 1628,74 | 4529,64 | 1114,86 |
| 66 | 2456623,65 | 4,74 | -76,24 | 100,9 | 2,7 | 57 | 55 | 169 | -55,20 | -915,42 | 1516,73 | -85,16 |
| 67 | 2456627,55 | 4,68 | -76,13 | 87,3 | 3,2 | 63 | 36 | 144 | 732,60 | 191,97 | 1742,15 | -1068,02 |
| 68 | 2456654,64 | 4,27 | -75,90 | 165,4 | 3,2 | 136 | 85 | 127 | 2363,97 | 535,60 | 3484,09 | -189,16 |
| 69 | 2456656,03 | 4,09 | -76,42 | 132,4 | 2,8 | 122 | 59 | 216 | 539,01 | -1247,94 | 3000,71 | 1166,08 |
| 70 | 2456341,30 | 4,81 | -76,27 | 95,8 | 2,5 | 115 | 88 | 141 | 64,36 | -521,30 | 713,62 | -922,43 |
| 71 | 2456349,53 | 4,65 | -76,30 | 110,5 | 2,7 | 102 | 72 | 160 | -2798,76 | 380,72 | 6372,43 | -1533,07 |
| 72 | 2456374,84 | 5,07 | -75,70 | 104,7 | 3,2 | -133 | 53 | 123 | -1771,39 | 877,82 | 2642,49 | -1602,82 |
| 73 | 2456361,05 | 5,03 | -76,30 | 93,1 | 3,2 | 273 | 63 | 218 | 2449,88 | 107,57 | 5092,67 | 1429,87 |
| 74 | 2456444,78 | 4,01 | -76,24 | 108,1 | 3,7 | 10 | 77 | 222 | -3559,02 | -490,94 | 2460,74 | 819,91 |
| 75 | 2456447,26 | 4,72 | -76,16 | 122,3 | 2,3 | 21 | 60 | 218 | 88,83 | -589,12 | 2428,50 | 788,05 |
| 76 | 2456493,92 | 3,96 | -76,28 | 138,1 | 2,7 | 132 | 71 | 188 | -3193,12 | -533,41 | 2914,08 | -436,62 |
| 77 | 2456494,87 | 5,13 | -75,47 | 115,3 | 3,5 | -54 | 85 | 140 | -3877,77 | 1485,25 | 4037,21 | -979,03 |
| 78 | 2456502,79 | 4,37 | -76,31 | 124,9 | 3,2 | 129 | 59 | 198 | -9,40 | -1590,88 | 1177,57 | -1003,07 |
| 79 | 2456476,97 | 4,88 | -76,22 | 109,0 | 2,7 | 160 | 81 | 90 | -1506,79 | 238,65 | 693,24 | -109,80 |
| 80 | 2456481,40 | 4,77 | -76,24 | 112,3 | 2,9 | -78 | 90 | 162 | -5823,09 | 188,97 | 3875,97 | -306,62 |
| 81 | 2456530,40 | 4,42 | -76,72 | 65,3 | 2,8 | 264 | 48 | 191 | 931,86 | -96,12 | 2554,11 | 593,37 |
| 82 | 2456530,76 | 4,50 | -76,30 | 122,3 | 3,6 | 157 | 64 | 217 | -825,92 | -1822,80 | 583,02 | -801,61 |
| 83 | 2456531,88 | 4,27 | -76,20 | 134,0 | 2,7 | 357 | 42 | 249 | 382,82 | 310,50 | 86,61 | -33,72 |
| 84 | 2456509,44 | 4,87 | -76,13 | 113,8 | 3,9 | 199 | 70 | 540 | -3483,10 | 371,98 | 2851,37 | 348,52 |
| 85 | 2456701,90 | 4,13 | -76,35 | 122,6 | 3,0 | 98 | 48 | 192 | -3101,36 | -689,95 | 2746,47 | 290,64 |
| 86 | 2456719,16 | 4,09 | -75,87 | 171,4 | 3,3 | 227 | 36 | 540 | 947,81 | 1799,75 | 2422,02 | 962,06 |
| 87 | 2456741,51 | 3,78 | -76,41 | 135,7 | 3,1 | 174 | 70 | 230 | 3256,66 | 1180,99 | 1974,11 | 706,94 |
| 88 | 2457075,43 | 4,80 | -76,06 | 102,7 | 5,2 | 93 | 51 | -179 | -1108,16 | 453,84 | 4474,56 | -109,35 |
| 89 | 2457076,04 | 2,73 | -76,46 | 161,3 | 5,6 | 119 | 45 | -156 | -2185,75 | -303,14 | 3228,94 | 563,45 |

Table S4. Focal mechanism information for 89 intermediate-depth events from 1966 to 2015 in the Cauca seismic cluster. At maximum strain before earthquake = MSBE and at earthquake origin time = AEOT. The data combines information of Cortes and Angelier (2005) with information from the RSNC, CMT Catalog (Dziewonski et al., 1981; Ekström et al., 2012), Chang et al., 2019, Salcedo et al., 2001, and Tabares et al., 1999. NP2 corresponds to the strike (S), dip (D) and rake (R) of the second nodal plane. Date in days of the Julian calendar since the reference time 4713 B.C.

| No. | Date (days in JC) | Lat º | Lon º | Depth (km) | Ml | NP2 | | | AEOT | | MSBE | |
|---|---|---|---|---|---|---|---|---|---|---|---|---|
| | | | | | | S | D | R | $\sigma_N$ (Pa) | $\sigma_S$ (Pa) | $\sigma_N$ (Pa) | $\sigma_S$ (Pa) |
| 1 | 2439129,26 | 4,65 | -76,00 | 98,0 | 5,1 | 255,96 | 67,53 | -13,00 | -2479,42 | -437,01 | 3034,53 | 228,99 |
| 2 | 2441797,28 | 5,20 | -75,80 | 117,0 | 5,3 | 289,94 | 69,21 | 171,44 | 124,29 | -264,96 | 758,97 | -649,39 |
| 3 | 2441776,08 | 4,70 | -75,80 | 158,0 | 6,1 | 121,93 | 59,74 | -159,04 | 1051,31 | -1184,82 | 5239,53 | -238,22 |
| 4 | 2442515,58 | 4,85 | -75,71 | 139,0 | 5,1 | 332,43 | 76,50 | 38,24 | -3386,11 | 915,41 | 3981,02 | 50,12 |
| 5 | 2442918,20 | 4,50 | -75,80 | 166,0 | 5,8 | 291,82 | 42,58 | -24,04 | -619,89 | -342,87 | 1107,67 | 832,01 |
| 6 | 2444201,49 | 4,81 | -76,20 | 105,0 | 6,3 | 120,00 | 60,00 | 0,00 | -1383,43 | -2275,06 | 2170,52 | 1217,50 |
| 7 | 2444023,04 | 5,28 | -75,73 | 122,0 | 4,9 | 7,77 | 46,83 | -80,48 | -1537,60 | -1249,95 | 249,47 | 146,17 |
| 8 | 2444416,00 | 4,70 | -75,35 | 151,0 | 6,3 | 137,07 | 76,55 | 163,54 | -435,69 | -757,98 | 2160,55 | -847,79 |
| 9 | 2446518,91 | 4,59 | -75,63 | 163,0 | 5,0 | 327,66 | 65,08 | 17,69 | 2095,16 | -1313,28 | 2740,62 | 911,77 |
| 10 | 2447527,37 | 5,10 | -75,78 | 118,0 | 4,9 | 297,17 | 70,28 | 134,70 | 1042,32 | -684,77 | 1286,27 | -834,04 |
| 11 | 2448219,44 | 4,45 | -75,46 | 147,0 | 6,0 | 53,33 | 75,15 | -72,24 | 620,40 | 232,33 | 1428,46 | 135,62 |
| 12 | 2448850,29 | 5,15 | -75,58 | 107,0 | 5,6 | 87,63 | 69,26 | -97,49 | 4313,28 | 1604,95 | 5331,81 | 1984,66 |
| 13 | 2450337,77 | 4,59 | -76,90 | 118,2 | 5,3 | 60,57 | 62,82 | 45,03 | 1010,82 | 510,06 | 4002,66 | -2115,01 |
| 14 | 2450794,06 | 4,11 | -75,84 | 189,5 | 6,3 | 139,12 | 65,71 | 138,68 | 1103,50 | -1610,28 | 4145,53 | -1502,22 |
| 15 | 2450499,27 | 4,78 | -76,50 | 118,1 | 5,8 | 298,17 | 49,73 | 76,88 | 526,84 | -701,66 | 2156,58 | -1654,75 |
| 16 | 2452174,64 | 4,03 | -76,17 | 178,0 | 5,9 | 137,12 | 65,71 | 138,68 | -1169,69 | -135,35 | 2417,27 | -2036,73 |
| 17 | 2454722,90 | 4,94 | -75,48 | 139,8 | 5,7 | 72,57 | 78,24 | -84,43 | -4970,73 | -1033,62 | 5528,71 | 1098,99 |
| 18 | 2455184,10 | 3,08 | -76,30 | 169,2 | 5,1 | 14,42 | 84,25 | -44,28 | -3993,94 | -658,08 | 1957,63 | 43,10 |
| 19 | 2455208,28 | 4,29 | -75,76 | 171,4 | 4,7 | 223,24 | 35,79 | 13,77 | -970,77 | 1070,99 | 568,06 | -464,82 |
| 20 | 2455226,24 | 4,74 | -76,30 | 127,4 | 4,7 | 229,49 | 43,15 | -61,64 | 2210,56 | 2139,92 | 2678,51 | 2221,15 |
| 21 | 2455199,78 | 3,87 | -76,06 | 112,9 | 3,7 | 30,00 | 33,00 | 90,00 | 1488,54 | -2292,15 | 1218,15 | -1875,78 |
| 22 | 2455247,69 | 4,73 | -76,14 | 99,8 | 3,1 | 149,03 | 47,86 | 20,43 | -1389,06 | 899,72 | 1403,59 | 507,05 |
| 23 | 2455269,95 | 4,66 | -76,16 | 93,8 | 2,9 | 234,01 | 63,24 | 61,95 | -3223,16 | 1701,51 | 4608,53 | -2438,95 |
| 24 | 2455258,82 | 4,81 | -76,22 | 110,7 | 4,0 | 201,36 | 52,55 | -83,03 | 3391,74 | 2563,02 | 2350,75 | 1601,44 |
| 25 | 2455300,20 | 4,61 | -76,45 | 97,5 | 3,9 | 70,84 | 51,97 | 45,23 | 3902,71 | -2417,38 | 4064,96 | -2714,54 |
| 26 | 2455395,14 | 4,73 | -76,29 | 107,0 | 2,8 | 288,50 | 61,12 | 17,19 | -2103,24 | -108,21 | 3583,29 | -215,03 |
| 27 | 2455464,80 | 4,75 | -76,20 | 95,9 | 3,0 | 272,97 | 39,67 | -32,40 | 1428,91 | 940,68 | 2661,75 | 1843,21 |
| 28 | 2455572,80 | 4,19 | -76,46 | 113,9 | 3,9 | 337,84 | 52,56 | 19,02 | -834,02 | 1168,03 | 643,72 | 886,84 |
| 29 | 2455563,72 | 4,52 | -76,27 | 96,8 | 3,0 | 112,97 | 68,79 | 16,12 | 1500,20 | -785,83 | 3120,95 | -299,91 |
| 30 | 2455590,56 | 4,85 | -76,08 | 115,3 | 4,3 | 31,85 | 78,41 | -15,32 | 299,05 | -231,51 | 1505,06 | 90,85 |
| 31 | 2455592,80 | 4,03 | -76,49 | 70,6 | 4,6 | 23,04 | 39,52 | -81,44 | 663,38 | 875,91 | 1648,02 | 1969,54 |
| 32 | 2455865,76 | 4,79 | -76,14 | 115,0 | 3,5 | 298,20 | 22,34 | -10,57 | 365,92 | 788,79 | 397,23 | 649,56 |
| 33 | 2455842,20 | 3,87 | -76,10 | 156,2 | 4,2 | 239,23 | 35,45 | 49,09 | 929,61 | -578,15 | 1571,62 | -1685,87 |
| 34 | 2455876,19 | 4,73 | -76,34 | 101,6 | 3,7 | 18,12 | 81,34 | -75,32 | 1927,46 | 11,74 | 3339,04 | 340,15 |
| 35 | 2455879,60 | 4,74 | -76,16 | 82,8 | 2,9 | 211,62 | 70,81 | 58,94 | -3606,54 | 1055,44 | 1236,46 | -1209,95 |
| 36 | 2455889,47 | 5,10 | -76,40 | 78,0 | 3,9 | 252,77 | 59,38 | 18,68 | -3083,66 | -536,29 | 4380,01 | -1152,63 |
| 37 | 2455898,10 | 4,84 | -76,30 | 90,9 | 2,8 | 220,53 | 82,21 | 30,31 | 1122,32 | 1146,93 | 2441,81 | -437,22 |
| 38 | 2455925,36 | 4,79 | -76,35 | 93,4 | 3,1 | 17,02 | 59,83 | -44,12 | 1311,76 | 311,57 | 814,33 | -138,54 |
| 39 | 2455927,24 | 3,87 | -76,21 | 111,6 | 3,0 | 352,98 | 38,96 | 36,57 | -1061,94 | 1186,70 | 267,56 | -213,93 |
| 40 | 2455641,07 | 4,90 | -76,22 | 116,2 | 4,4 | 227,93 | 67,68 | -22,79 | -1787,35 | -2192,76 | 6562,40 | -328,94 |
| 41 | 2455672,79 | 4,14 | -76,23 | 140,5 | 3,8 | 0,25 | 58,42 | 32,20 | 769,95 | -308,14 | 1626,48 | -385,70 |
| 42 | 2455716,41 | 3,97 | -76,72 | 114,5 | 3,4 | 197,90 | 79,79 | 56,34 | 2530,72 | 164,46 | 2369,21 | -692,70 |
| 43 | 2455804,64 | 4,63 | -76,26 | 91,8 | 3,6 | 347,50 | 16,96 | 62,55 | -207,98 | 788,34 | 460,25 | -1273,34 |
| 44 | 2455824,16 | 4,31 | -76,33 | 118,4 | 3,5 | 0,00 | 90,00 | 44,00 | -1689,46 | -492,60 | 309,11 | 250,03 |
| 45 | 2455831,48 | 4,71 | -76,19 | 83,3 | 3,0 | 161,36 | 79,09 | 43,99 | -4352,45 | 1028,68 | 5999,04 | -349,49 |
| 46 | 2455958,19 | 4,73 | -76,22 | 76,6 | 2,9 | 146,49 | 30,75 | -49,83 | -292,36 | -459,70 | -133,96 | -4,31 |
| 47 | 2455931,00 | 4,01 | -76,72 | 100,7 | 3,1 | 192,00 | 90,00 | 72,00 | -2000,71 | -338,81 | 2232,73 | 63,56 |
| 48 | 2456222,48 | 4,48 | -75,93 | 139,4 | 3,8 | 221,46 | 59,63 | -12,78 | 1542,25 | 906,53 | 1480,20 | 171,68 |
| 49 | 2456207,51 | 4,78 | -76,26 | 110,2 | 3,8 | 50,14 | 74,07 | -41,95 | -961,10 | 856,70 | 774,02 | -706,55 |
| 50 | 2455977,49 | 4,74 | -76,18 | 88,7 | 3,6 | 207,96 | 82,96 | 57,68 | -4220,13 | 208,01 | 4152,71 | -805,23 |
| 51 | 2455981,93 | 3,88 | -76,34 | 107,6 | 3,3 | 78,09 | 59,55 | -29,36 | -2572,50 | 465,24 | 4816,73 | 638,99 |
| 52 | 2455987,57 | 4,37 | -76,53 | 96,9 | 4,5 | 231,89 | 81,20 | 12,15 | 341,63 | 1811,43 | 480,05 | -158,24 |
| 53 | 2456009,37 | 4,71 | -76,20 | 86,1 | 2,5 | 260,70 | 70,28 | 71,17 | -2949,97 | 1246,86 | 5990,42 | -2143,65 |
| 54 | 2456037,69 | 4,75 | -76,22 | 92,0 | 2,7 | 64,26 | 69,01 | -18,25 | 4466,18 | -251,11 | 4599,24 | -505,80 |
| 55 | 2456088,62 | 4,10 | -76,64 | 68,6 | 3,2 | 193,00 | 90,00 | 57,00 | -555,64 | 69,25 | 1829,19 | -408,27 |
| 56 | 2456105,66 | 3,76 | -76,30 | 143,4 | 4,5 | 238,99 | 78,94 | 50,26 | -1981,42 | 725,71 | 3573,73 | -721,05 |
| 57 | 2456149,61 | 3,88 | -76,13 | 153,7 | 3,0 | 153,29 | 48,57 | -18,82 | -156,00 | -769,94 | -246,09 | 303,42 |
| 58 | 2456147,23 | 4,05 | -76,15 | 145,3 | 2,8 | 317,00 | 90,00 | 15,00 | -723,23 | 1894,42 | 2854,33 | 1558,45 |
| 59 | 2456148,67 | 4,75 | -76,17 | 91,9 | 2,5 | 149,17 | 37,65 | -35,92 | -654,84 | -468,46 | -31,31 | 427,68 |
| 60 | 2456306,83 | 4,97 | -76,02 | 113,8 | 3,0 | 1,15 | 60,02 | -53,32 | 4229,70 | 1818,61 | 2864,74 | 1155,45 |
| 61 | 2456308,58 | 4,48 | -76,20 | 130,4 | 3,0 | 319,95 | 81,05 | 6,07 | -2271,05 | -1404,91 | 3488,31 | 1734,31 |
| 62 | 2456295,20 | 4,15 | -76,40 | 129,2 | 4,4 | 61,81 | 34,46 | 10,64 | 57,51 | -1056,97 | 1404,61 | -1067,93 |
| 63 | 2456584,52 | 4,90 | -75,90 | 118,3 | 3,0 | 80,98 | 63,02 | -2,24 | -2145,81 | -663,11 | 5262,49 | -119,14 |
| 64 | 2456573,60 | 4,23 | -76,51 | 60,8 | 3,6 | 202,92 | 36,96 | 77,85 | -1679,99 | 2145,06 | 819,32 | -1298,04 |
| 65 | 2456614,14 | 4,37 | -75,92 | 156,7 | 3,7 | 149,75 | 67,38 | -20,65 | 216,93 | 1628,85 | 3390,17 | 1114,89 |
| 66 | 2456623,65 | 4,74 | -76,24 | 100,9 | 2,7 | 153,36 | 81,01 | 35,50 | 1655,75 | -915,44 | 2595,63 | -85,11 |
| 67 | 2456627,55 | 4,68 | -76,13 | 87,3 | 3,2 | 183,45 | 69,79 | 59,55 | -308,59 | 191,87 | 2656,10 | -1068,03 |
| 68 | 2456654,64 | 4,27 | -75,90 | 165,4 | 3,2 | 232,60 | 37,29 | 8,27 | 940,08 | 535,66 | 1190,11 | -189,13 |
| 69 | 2456656,03 | 4,09 | -76,42 | 132,4 | 2,8 | 11,48 | 59,75 | -36,60 | -917,99 | -1247,84 | 3259,97 | 1165,94 |
| 70 | 2456341,30 | 4,81 | -76,27 | 95,8 | 2,5 | 206,62 | 51,03 | 2,57 | -1539,78 | -521,42 | -1,86 | -922,46 |
| 71 | 2456349,53 | 4,65 | -76,30 | 110,5 | 2,7 | 198,42 | 71,02 | 19,07 | -3842,82 | 380,53 | 3649,68 | -1533,11 |
| 72 | 2456374,84 | 5,07 | -75,70 | 104,7 | 3,2 | 359,82 | 47,95 | 54,14 | -1358,70 | 877,77 | 2071,92 | -1602,77 |

| No. | Date (days in JC) | Lat º | Lon º | Depth (km) | Ml | NP2 | | | AEOT | | MSBE | |
|---|---|---|---|---|---|---|---|---|---|---|---|---|
| | | | | | | S | D | R | $\sigma_N$ (Pa) | $\sigma_S$ (Pa) | $\sigma_N$ (Pa) | $\sigma_S$ (Pa) |
| 73 | 2456361,05 | 5,03 | -76,30 | 93,1 | 3,2 | 163,47 | 56,73 | -32,89 | -1172,13 | 107,72 | 3684,87 | 1430,03 |
| 74 | 2456444,78 | 4,01 | -76,24 | 108,1 | 3,7 | 268,55 | 49,31 | -17,26 | -543,18 | -490,95 | 1905,83 | 819,96 |
| 75 | 2456447,26 | 4,72 | -76,16 | 122,3 | 2,3 | 269,66 | 57,78 | -36,23 | -2086,07 | -589,04 | 2322,62 | 788,06 |
| 76 | 2456493,92 | 3,96 | -76,28 | 138,1 | 2,7 | 39,38 | 82,44 | -19,17 | -6484,78 | -533,30 | 3696,37 | -436,70 |
| 77 | 2456494,87 | 5,13 | -75,47 | 115,3 | 3,5 | 40,18 | 50,18 | 6,52 | -3527,93 | 1485,51 | 2574,37 | -979,18 |
| 78 | 2456502,79 | 4,37 | -76,31 | 124,9 | 3,2 | 29,50 | 74,64 | -32,28 | -2551,49 | -1590,90 | 1147,03 | -1003,14 |
| 79 | 2456476,97 | 4,88 | -76,22 | 109,0 | 2,7 | 340,00 | 9,00 | 90,00 | -37,80 | 238,65 | 17,39 | -109,80 |
| 80 | 2456481,40 | 4,77 | -76,24 | 112,3 | 2,9 | 12,00 | 72,00 | 0,00 | -1315,43 | 188,97 | 2683,04 | -306,62 |
| 81 | 2456530,40 | 4,42 | -76,72 | 65,3 | 2,8 | 166,59 | 81,85 | -42,53 | 3219,16 | -96,16 | 251,24 | 593,32 |
| 82 | 2456530,76 | 4,50 | -76,30 | 122,3 | 3,6 | 48,72 | 57,25 | -31,41 | 273,82 | -1822,75 | 2090,51 | -801,58 |
| 83 | 2456531,88 | 4,27 | -76,20 | 134,0 | 2,7 | 204,32 | 51,34 | -72,12 | 519,67 | 310,53 | 153,84 | -33,68 |
| 84 | 2456509,44 | 4,87 | -76,13 | 113,8 | 3,9 | 289,00 | 90,00 | 20,00 | -5303,10 | 371,98 | 3871,20 | 348,52 |
| 85 | 2456701,90 | 4,13 | -76,35 | 122,6 | 3,0 | 359,91 | 81,11 | -42,63 | -2192,73 | -689,78 | 3461,94 | 290,62 |
| 86 | 2456719,16 | 4,09 | -75,87 | 171,4 | 3,3 | 317,00 | 90,00 | 54,00 | 2719,57 | 1799,75 | 6627,20 | 962,06 |
| 87 | 2456741,51 | 3,78 | -76,41 | 135,7 | 3,1 | 61,82 | 43,96 | -29,52 | 1621,24 | 1181,01 | 1163,82 | 706,93 |
| 88 | 2457075,43 | 4,80 | -76,06 | 102,7 | 5,2 | 2,37 | 89,22 | -39,00 | 4218,39 | 453,94 | 4418,98 | -109,20 |
| 89 | 2457076,04 | 2,73 | -76,46 | 161,3 | 5,6 | 11,52 | 73,29 | -47,59 | -1008,62 | -303,29 | 2996,43 | 563,51 |

Table S5. Focal mechanism information for intermediate-depth events from 1982 to 2016 in the region of the El Puyo seismic cluster. At maximum strain before earthquake = MSBE and at earthquake origin time = AEOT. The information about focal mechanism was obtained from CMT Catalog (Dziewonski et al., 1981; Ekström et al., 2012). Conventions as in Tables S1.

| No. | Date (days in JC) | Lat º | Lon º | Depth (km) | Ml | NP1 | | | AEOT | | MSBE | |
|---|---|---|---|---|---|---|---|---|---|---|---|---|
| | | | | | | S | D | R | $\sigma_N$ (Pa) | $\sigma_S$ (Pa) | $\sigma_N$ (Pa) | $\sigma_S$ (Pa) |
| 1 | 2445110,01 | -1,41 | -78,25 | 181,1 | 5,3 | 176 | 19 | -36 | -54,27 | 176,71 | 481,37 | 763,31 |
| 2 | 2446451,79 | -1,84 | -77,76 | 166,3 | 5,4 | 157 | 25 | -51 | -467,72 | -1048,07 | 191,74 | 314,07 |
| 3 | 2447147,71 | -1,26 | -77,90 | 172,9 | 5,5 | 142 | 25 | -76 | 230,09 | 457,26 | 595,66 | 1288,31 |
| 4 | 2447420,28 | -1,25 | -78,01 | 169,0 | 6,2 | 121 | 28 | -109 | 804,61 | 1355,04 | 977,72 | 1682,32 |
| 5 | 2449377,74 | -1,72 | -77,79 | 154,3 | 5,4 | 195 | 35 | -36 | 800,74 | 302,58 | 987,32 | 651,44 |
| 6 | 2450110,52 | -1,79 | -77,71 | 149,0 | 5,3 | 149 | 24 | -63 | 161,95 | 305,73 | 223,68 | 509,17 |
| 7 | 2451295,26 | -1,50 | -77,83 | 164,2 | 6,0 | 169 | 33 | -49 | 210,10 | 88,95 | 1113,48 | 1337,37 |
| 8 | 2451419,03 | -1,36 | -77,75 | 198,2 | 6,2 | 122 | 18 | -107 | -372,75 | -1036,30 | 604,65 | 1709,42 |
| 9 | 2453728,41 | -1,59 | -77,76 | 196,8 | 6,1 | 120 | 21 | -107 | 285,29 | 712,27 | 232,37 | 525,26 |
| 10 | 2453542,92 | -1,86 | -78,02 | 166,1 | 5,0 | 131 | 16 | -94 | -326,61 | -1114,29 | 282,14 | 971,92 |
| 11 | 2454039,91 | -1,53 | -78,04 | 177,5 | 5,4 | 137 | 29 | -80 | 115,61 | 237,08 | 870,62 | 1576,54 |
| 12 | 2453908,03 | -2,06 | -77,42 | 165,8 | 5,3 | 185 | 36 | -29 | 533,01 | 622,78 | 1214,07 | 554,54 |
| 13 | 2454106,77 | -1,52 | -78,09 | 164,4 | 5,0 | 127 | 40 | -102 | 1524,66 | 1694,22 | 1605,11 | 1800,70 |
| 14 | 2454145,12 | -1,60 | -78,05 | 175,8 | 5,6 | 139 | 25 | -95 | 381,66 | 809,85 | 160,08 | 340,08 |
| 15 | 2454155,09 | -1,58 | -78,08 | 160,1 | 5,0 | 95 | 52 | -154 | -918,17 | -113,14 | 1826,63 | 392,94 |
| 16 | 2454188,37 | -1,60 | -77,81 | 191,5 | 5,3 | 104 | 24 | -124 | -504,94 | -720,38 | 599,85 | 1070,12 |
| 17 | 2454303,29 | -1,66 | -78,30 | 158,7 | 5,2 | 132 | 23 | -101 | -67,56 | -173,19 | 377,78 | 850,61 |
| 18 | 2454651,75 | -1,58 | -77,91 | 193,4 | 5,2 | 122 | 32 | -119 | 1342,08 | 1740,18 | 1135,36 | 1412,63 |
| 19 | 2454952,52 | -1,74 | -77,66 | 194,9 | 5,1 | 153 | 21 | -62 | 8,76 | -128,54 | 200,82 | 464,41 |
| 20 | 2455421,00 | -1,51 | -77,51 | 197,8 | 7,1 | 153 | 21 | -68 | -398,03 | -1045,80 | 810,97 | 2070,14 |
| 21 | 2456070,93 | -1,96 | -78,00 | 158,8 | 4,8 | 125 | 15 | -102 | -156,62 | -523,68 | 203,10 | 719,60 |
| 22 | 2456973,08 | -2,00 | -78,23 | 163,2 | 4,8 | 135 | 37 | -110 | -1450,77 | -1701,70 | 757,35 | 853,28 |
| 23 | 2457662,20 | -1,80 | -77,75 | 185,0 | 5,0 | 159 | 24 | -71 | 770,36 | 1717,20 | 777,33 | 1725,25 |

Table S6. Focal mechanism information for intermediate-depth events from 1982 to 2016 in the region of the El Puyo seismic cluster. At maximum strain before earthquake = MSBE and at earthquake origin time = AEOT. The information about focal mechanism was obtained from CMT Catalog (Dziewonski et al., 1981; Ekström et al., 2012). Conventions as in Table S2.

| No. | Date (days in JC) | Lat º | Lon º | Depth (km) | Ml | NP2 | | | AEOT | | MSBE | |
|---|---|---|---|---|---|---|---|---|---|---|---|---|
| | | | | | | S | D | R | $\sigma_N$ (Pa) | $\sigma_S$ (Pa) | $\sigma_N$ (Pa) | $\sigma_S$ (Pa) |
| 1 | 2445110,01 | -1,41 | -78,25 | 181,1 | 5,3 | 300.49 | 78.97 | -105.57 | 757,65 | 176,67 | 5266,47 | 763,02 |
| 2 | 2446451,79 | -1,84 | -77,76 | 166,3 | 5,4 | 295.22 | 70.83 | -106.35 | -2938,38 | -1047,90 | 1138,92 | 314,03 |
| 3 | 2447147,71 | -1,26 | -77,90 | 172,9 | 5,5 | 306.62 | 65.79 | -96.44 | 1096,77 | 457,25 | 2999,05 | 1288,31 |
| 4 | 2447420,28 | -1,25 | -78,01 | 169,0 | 6,2 | 322.3 | 63.65 | -80.18 | 2577,99 | 1354,99 | 3187,65 | 1682,25 |
| 5 | 2449377,74 | -1,72 | -77,79 | 154,3 | 5,4 | 315.76 | 70.3 | -119.53 | 997,51 | 302,56 | 2879,98 | 651,34 |
| 6 | 2450110,52 | -1,79 | -77,71 | 149,0 | 5,3 | 299.85 | 68.75 | -101.43 | 861,14 | 305,75 | 1340,35 | 509,20 |
| 7 | 2451295,26 | -1,50 | -77,83 | 164,2 | 6,0 | 302.97 | 65.73 | -113.08 | 128,47 | 88,93 | 3685,62 | 1337,30 |
| 8 | 2451419,03 | -1,36 | -77,75 | 198,2 | 6,2 | 319.82 | 72.81 | -84.57 | -3258,23 | -1036,40 | 5291,97 | 1709,60 |
| 9 | 2453728,41 | -1,59 | -77,76 | 196,8 | 6,1 | 318.13 | 69.96 | -83.6 | 1849,49 | 712,20 | 1368,72 | 525,22 |
| 10 | 2453542,92 | -1,86 | -78,02 | 166,1 | 5,0 | 315.16 | 74.04 | -88.85 | -3834,17 | -1114,36 | 3366,88 | 971,96 |
| 11 | 2454039,91 | -1,53 | -78,04 | 177,5 | 5,4 | 305.6 | 61.48 | -95.5 | 464,06 | 237,09 | 2967,90 | 1576,59 |
| 12 | 2453908,03 | -2,06 | -77,42 | 165,8 | 5,3 | 299.15 | 73.44 | -122.43 | 2743,19 | 622,99 | 3445,35 | 554,78 |
| 13 | 2454106,77 | -1,52 | -78,09 | 164,4 | 5,0 | 322.51 | 51.04 | -80.1 | 2004,26 | 1694,27 | 2162,54 | 1800,72 |
| 14 | 2454145,12 | -1,60 | -78,05 | 175,8 | 5,6 | 324.51 | 65.1 | -87.67 | 1736,29 | 809,90 | 737,29 | 340,10 |
| 15 | 2454155,09 | -1,58 | -78,08 | 160,1 | 5,0 | 348.29 | 69.79 | -41 | -524,99 | -113,10 | 1871,57 | 392,93 |

| No. | Date (days in JC) | Lat º | Lon º | Depth (km) | Ml | NP2 | | | AEOT | | MSBE | |
|---|---|---|---|---|---|---|---|---|---|---|---|---|
| | | | | | | S | D | R | $\sigma_N$ (Pa) | $\sigma_S$ (Pa) | $\sigma_N$ (Pa) | $\sigma_S$ (Pa) |
| 16 | 2454188,37 | -1,60 | -77,81 | 191,5 | 5,3 | 320.44 | 70.29 | -76.02 | -1945,80 | -720,49 | 2838,23 | 1070,27 |
| 17 | 2454303,29 | -1,66 | -78,30 | 158,7 | 5,2 | 323.92 | 67.45 | -85.37 | -456,77 | -173,16 | 1954,85 | 850,52 |
| 18 | 2454651,75 | -1,58 | -77,91 | 193,4 | 5,2 | 335.17 | 62.39 | -73.15 | 3300,15 | 1740,12 | 2696,56 | 1412,58 |
| 19 | 2454952,52 | -1,74 | -77,66 | 194,9 | 5,1 | 303.34 | 71.55 | -100.22 | -189,40 | -128,56 | 1637,15 | 464,45 |
| 20 | 2455421,00 | -1,51 | -77,51 | 197,8 | 7,1 | 309.6 | 70.59 | -98.18 | -3277,26 | -1045,99 | 6313,55 | 2070,50 |
| 21 | 2456070,93 | -1,96 | -78,00 | 158,8 | 4,8 | 317.41 | 75.34 | -86.81 | -1846,04 | -523,56 | 2656,50 | 719,41 |
| 22 | 2456973,08 | -2,00 | -78,23 | 163,2 | 4,8 | 339.5 | 55.56 | -75.55 | -2259,92 | -1701,72 | 1186,04 | 853,30 |
| 23 | 2457662,20 | -1,80 | -77,75 | 185,0 | 5,0 | 318.35 | 67.38 | -98.25 | 4444,98 | 1717,30 | 4451,86 | 1725,35 |

Table S7. Focal mechanism information for intermediate-depth events from 1977 to 2017 in the region of the Pucallpa seismic cluster. At maximum strain before earthquake = MSBE and at earthquake origin time = AEOT. The information about focal mechanism was obtained from CMT Catalog (Dziewonski et al., 1981; Ekström et al., 2012). Conventions as in Table S1.

| No. | Date (days in JC) | Lat º | Lon º | Depth (km) | Ml | NP1 | | | AEOT | | MSBE | |
|---|---|---|---|---|---|---|---|---|---|---|---|---|
| | | | | | | S | D | R | $\sigma_N$ (Pa) | $\sigma_S$ (Pa) | $\sigma_N$ (Pa) | $\sigma_S$ (Pa) |
| 1 | 2443216,38 | -8,0 | -74,5 | 153,0 | 5,9 | 62 | 17 | 149 | -166,59 | 422,62 | 267,06 | -424,40 |
| 2 | 2444333,77 | -7,4 | -74,5 | 164,7 | 5,1 | 196 | 59 | -17 | 2773,39 | 199,34 | 2300,94 | 121,78 |
| 3 | 2445304,42 | -8,5 | -73,5 | 149,7 | 5,4 | 291 | 52 | 6 | -3427,85 | -899,80 | 4183,75 | 102,83 |
| 4 | 2446529,25 | -7,8 | -73,8 | 179,0 | 5,6 | 206 | 27 | -66 | 884,15 | 1461,58 | 1049,47 | 1739,64 |
| 5 | 2447545,21 | -7,3 | -74,3 | 148,1 | 5,4 | 10 | 47 | -117 | 930,12 | 636,85 | 1728,44 | 1489,12 |
| 6 | 2447864,09 | -7,4 | -74,3 | 150,9 | 6,5 | 192 | 40 | -70 | -713,81 | -952,80 | 582,14 | 584,40 |
| 7 | 2448586,09 | -8,5 | -74,5 | 146,6 | 5,3 | 211 | 11 | -41 | -76,29 | -259,63 | 76,34 | 159,86 |
| 8 | 2449728,93 | -7,9 | -74,2 | 177,0 | 5,4 | 157 | 39 | -164 | -407,53 | 129,32 | 230,31 | -58,25 |
| 9 | 2450198,21 | -7,9 | -74,0 | 160,6 | 5,6 | 222 | 55 | -39 | -1600,68 | -583,00 | 1004,18 | 407,87 |
| 10 | 2450883,24 | -8,1 | -73,9 | 161,9 | 5,2 | 221 | 27 | -65 | 625,56 | 1073,87 | 985,51 | 1631,68 |
| 11 | 2450907,42 | -8,0 | -74,5 | 153,7 | 6,6 | 68 | 13 | 166 | 38,73 | 36,93 | 115,22 | -106,76 |
| 12 | 2451849,69 | -8,0 | -74,4 | 153,1 | 5,9 | 75 | 10 | -179 | -42,49 | 145,66 | 88,58 | 18,91 |
| 13 | 2452272,42 | -8,2 | -74,2 | 161,5 | 5,9 | 303 | 4 | 50 | -22,10 | 184,14 | 23,26 | -235,33 |
| 14 | 2452123,11 | -7,2 | -74,0 | 195,4 | 5,3 | 20 | 43 | -53 | 377,25 | 286,35 | 1454,18 | 1153,33 |
| 15 | 2453151,66 | -8,4 | -74,4 | 164,1 | 5,2 | 32 | 8 | 128 | -17,27 | 99,92 | 16,59 | -54,85 |
| 16 | 2453893,60 | -7,6 | -73,9 | 197,2 | 5,2 | 234 | 45 | -61 | 1895,36 | 1578,39 | 1980,18 | 1637,75 |
| 17 | 2454293,72 | -8,0 | -74,4 | 156,4 | 6,1 | 215 | 37 | -29 | 573,74 | 551,39 | 1482,27 | 828,84 |
| 18 | 2454705,38 | -7,8 | -74,5 | 158,2 | 6,4 | 198 | 39 | -60 | -858,77 | -1055,62 | 552,25 | 409,39 |
| 19 | 2455222,45 | -8,5 | -74,5 | 148,3 | 5,9 | 83 | 33 | -130 | -430,89 | -126,71 | 831,40 | 902,47 |
| 20 | 2455464,29 | -7,9 | -74,5 | 155,8 | 5,5 | 207 | 27 | -38 | 724,39 | 748,64 | 1040,86 | 1006,09 |
| 21 | 2455717,07 | -7,8 | -74,5 | 154,2 | 5,0 | 163 | 43 | -100 | -1418,03 | -1595,61 | 1800,18 | 1887,08 |
| 22 | 2455798,24 | -7,7 | -74,7 | 144,4 | 7,0 | 197 | 40 | -57 | -468,17 | -675,02 | 478,55 | 352,65 |
| 23 | 2456141,90 | -8,6 | -74,4 | 150,5 | 6,0 | 222 | 26 | -66 | -740,83 | -1227,90 | 1189,93 | 2093,06 |
| 24 | 2457158,55 | -7,8 | -74,4 | 160,7 | 5,0 | 8 | 42 | -72 | 1176,38 | 1172,75 | 1802,31 | 1850,94 |
| 25 | 2457418,82 | -8,3 | -74,3 | 160,7 | 5,3 | 82 | 15 | 179 | 77,11 | -36,02 | 127,93 | 107,56 |
| 26 | 2457704,12 | -8,4 | -74,5 | 148,4 | 4,9 | 140 | 48 | -170 | 2954,05 | -87,44 | 2666,95 | -183,30 |
| 27 | 2457448,87 | -7,9 | -74,4 | 155,6 | 5,0 | 208 | 38 | -46 | 613,75 | 448,18 | 316,75 | 263,25 |
| 28 | 2457877,60 | -8,6 | -74,4 | 156,1 | 5,3 | 73 | 22 | -167 | 272,92 | -58,51 | 396,81 | 156,43 |

Table S8. Focal mechanism information for intermediate-depth events from 1977 to 2017 in the region of the Pucallpa seismic cluster. At maximum strain before earthquake = MSBE and at earthquake origin time = AEOT. The information about focal mechanism was obtained from CMT Catalog (Dziewonski et al., 1981; Ekström et al., 2012). Conventions as in Table S2.

| No. | Date (days in JC) | Lat º | Lon º | Depth (km) | Ml | NP2 | | | AEOT | | MSBE | |
|---|---|---|---|---|---|---|---|---|---|---|---|---|
| | | | | | | S | D | R | $\sigma_N$ (Pa) | $\sigma_S$ (Pa) | $\sigma_N$ (Pa) | $\sigma_S$ (Pa) |
| 1 | 2443216,38 | -8,0 | -74,5 | 153,0 | 5,9 | 181,9 | 81,3 | 75,3 | -1351,81 | 422,53 | 2644,65 | -424,36 |
| 2 | 2444333,77 | -7,4 | -74,5 | 164,7 | 5,1 | 295,0 | 75,5 | -147,9 | 4285,53 | 199,14 | 3979,81 | 121,58 |
| 3 | 2445304,42 | -8,5 | -73,5 | 149,7 | 5,4 | 197,3 | 85,3 | 141,9 | -5432,56 | -900,15 | 5822,68 | 103,20 |
| 4 | 2446529,25 | -7,8 | -73,8 | 179,0 | 5,6 | 359,5 | 65,5 | -101,7 | 3153,58 | 1461,46 | 3907,37 | 1739,48 |
| 5 | 2447545,21 | -7,3 | -74,3 | 148,1 | 5,4 | 226,8 | 49,3 | -64,0 | 761,03 | 636,88 | 1990,12 | 1489,13 |
| 6 | 2447864,09 | -7,4 | -74,3 | 150,9 | 6,5 | 346,6 | 52,8 | -106,0 | -1614,47 | -952,81 | 852,13 | 584,43 |
| 7 | 2448586,09 | -8,5 | -74,5 | 146,6 | 5,3 | 341,5 | 82,8 | -98,4 | -3316,28 | -259,51 | 1879,91 | 159,78 |
| 8 | 2449728,93 | -7,9 | -74,2 | 177,0 | 5,4 | 54,4 | 80,0 | -52,1 | -371,28 | 129,34 | 750,63 | -58,24 |
| 9 | 2450198,21 | -7,9 | -74,0 | 160,6 | 5,6 | 336,9 | 59,0 | -138,0 | -854,22 | -582,94 | 1167,80 | 407,87 |
| 10 | 2450883,24 | -8,1 | -73,9 | 161,9 | 5,2 | 13,4 | 65,7 | -102,2 | 2374,89 | 1073,98 | 3545,50 | 1631,82 |
| 11 | 2450907,42 | -8,0 | -74,5 | 153,7 | 6,6 | 171,7 | 86,9 | 77,4 | 348,78 | 36,95 | 2016,28 | -106,76 |
| 12 | 2451849,69 | -8,0 | -74,4 | 153,1 | 5,9 | 344,9 | 89,8 | -80,0 | -679,92 | 145,74 | 2558,75 | 18,75 |
| 13 | 2452272,42 | -8,2 | -74,2 | 161,5 | 5,9 | 163,1 | 86,9 | 92,6 | -4001,62 | 183,97 | 4316,89 | -235,09 |
| 14 | 2452123,11 | -7,2 | -74,0 | 195,4 | 5,3 | 154,1 | 57,0 | -119,3 | 960,40 | 286,37 | 2121,61 | 1153,31 |
| 15 | 2453151,66 | -8,4 | -74,4 | 164,1 | 5,2 | 173,7 | 83,7 | 85,1 | -1557,28 | 99,95 | 247,04 | -54,90 |
| 16 | 2453893,60 | -7,6 | -73,9 | 197,2 | 5,2 | 15,9 | 51,8 | -115,9 | 2015,90 | 1578,42 | 2074,80 | 1637,78 |
| 17 | 2454293,72 | -8,0 | -74,4 | 156,4 | 6,1 | 328,9 | 73,0 | -123,4 | 2079,94 | 551,27 | 3431,85 | 828,66 |
| 18 | 2454705,38 | -7,8 | -74,5 | 158,2 | 6,4 | 341,4 | 57,0 | -112,0 | -2299,16 | -1055,83 | 714,72 | 409,42 |

| No. | Date (days in JC) | Lat º | Lon º | Depth (km) | Ml | NP2 | | | AEOT | | MSBE | |
|---|---|---|---|---|---|---|---|---|---|---|---|---|
| | | | | | | S | D | R | $\sigma_N$ (Pa) | $\sigma_S$ (Pa) | $\sigma_N$ (Pa) | $\sigma_S$ (Pa) |
| 19 | 2455222,45 | -8,5 | -74,5 | 148,3 | 5,9 | 308,0 | 65,3 | -67,3 | -174,27 | -126,79 | 2035,90 | 902,50 |
| 20 | 2455464,29 | -7,9 | -74,5 | 155,8 | 5,5 | 331,8 | 73,8 | -111,9 | 3021,15 | 748,57 | 4606,91 | 1005,97 |
| 21 | 2455717,07 | -7,8 | -74,5 | 154,2 | 5,0 | 356,6 | 47,8 | -80,8 | -1949,09 | -1595,64 | 2114,65 | 1887,04 |
| 22 | 2455798,24 | -7,7 | -74,7 | 144,4 | 7,0 | 336,7 | 57,4 | -114,6 | -1644,48 | -675,02 | 737,10 | 352,64 |
| 23 | 2456141,90 | -8,6 | -74,4 | 150,5 | 6,0 | 15,7 | 66,4 | -101,2 | -2469,36 | -1228,00 | 4657,72 | 2093,23 |
| 24 | 2457158,55 | -7,8 | -74,4 | 160,7 | 5,0 | 164,4 | 50,5 | -105,6 | 1510,68 | 1172,72 | 2453,75 | 1850,89 |
| 25 | 2457418,82 | -8,3 | -74,3 | 160,7 | 5,3 | 173,0 | 89,7 | 75,0 | 624,22 | -36,04 | 1544,14 | 107,54 |
| 26 | 2457704,12 | -8,4 | -74,5 | 148,4 | 4,9 | 43,3 | 82,6 | -42,4 | 6034,61 | -87,69 | 4671,32 | -183,48 |
| 27 | 2457448,87 | -7,9 | -74,4 | 155,6 | 5,0 | 337,2 | 63,7 | -118,5 | 962,93 | 448,18 | 1037,93 | 263,32 |
| 28 | 2457877,60 | -8,6 | -74,4 | 156,1 | 5,3 | 330,9 | 85,2 | -68,5 | 545,97 | -58,57 | 1436,08 | 156,32 |